\documentclass{IEEEoj}

\usepackage{cite}
\usepackage{amsmath,amssymb,amsfonts}
\usepackage{graphicx,color}
\usepackage{textcomp}

\usepackage{mathtools}
\usepackage{booktabs}
\usepackage{multirow}
\usepackage{array}
\usepackage{algorithm}
\usepackage{algpseudocode}
\usepackage{xcolor}
\usepackage{url}
\usepackage{acro}
\usepackage{comment}

\AtBeginDocument{\definecolor{ojcolor}{cmyk}{0.93,0.59,0.15,0.02}}
\def\OJlogo{\vspace{-4pt}\hskip-4pt\includegraphics[height=18pt]{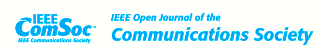}}

\DeclareAcronym{wtr}{
  short = WTR,
  long  = Wireless Technology Recognition
}

\DeclareAcronym{sdr}{
    short = SDR,
    long = software-defined radio
}

\DeclareAcronym{iq}{
    short = IQ,
    long = In-phase and Quadrature
}

\DeclareAcronym{cfo}{
    short = CFO,
    long = Carrier Frequency Offset
}

\DeclareAcronym{awgn}{
    short = AWGN,
    long = Additive White Gaussian Noise
}

\DeclareAcronym{mlp}{
    short = MLP,
    long = Multilayer Perceptron
}

\DeclareAcronym{relu}{
    short = ReLU,
    long = Rectified Linear Unit
}

\DeclareAcronym{snr}{
    short = SNR,
    long = Signal-to-Noise Ratio
}

\DeclareAcronym{lna}{
    short = LNA,
    long = Low Noise Amplifier
}

\DeclareAcronym{stft}{
    short = STFT,
    long = Short-Time Fourier Transform
}

\DeclareAcronym{iot}{
    short = IoT,
    long = Internet-of-Things
}

\DeclareAcronym{its}{
    short = ITS,
    long = Intelligent Transportation System
}

\DeclareAcronym{ml}{
    short = ML,
    long = Machine Learning
}

\DeclareAcronym{dl}{
    short = DL,
    long = Deep Learning
}

\DeclareAcronym{fft}{
    short = FFT,
    long = Fast Fourier Transform 
}

\DeclareAcronym{cnn}{
    short = CNN,
    long = Convolutional Neural Network
}

\DeclareAcronym{wfm}{
    short = WFM,
    long = Wireless Foundation Model
}

\DeclareAcronym{dnn}{
    short = DNN,
    long = Deep Neural Network
}

\DeclareAcronym{svm}{
    short = SVM,
    long = Support Vector Machine
}

\DeclareAcronym{rssi}{
    short = RSSI,
    long = Received Signal Strength Indicator
}

\DeclareAcronym{psd}{
    short = PSD,
    long = Power Spectral Density
}

\DeclareAcronym{scf}{
    short = SCF,
    long = Spectral Correlation Function
}

\DeclareAcronym{acf}{
    short = ACF,
    long = Auto-Correlation Function
}

\DeclareAcronym{lte}{
    short = LTE,
    long = Long Term Evolution
}

\DeclareAcronym{5g}{
    short = 5G NR,
    long = 5G New Radio
}

\DeclareAcronym{qos}{
    short = QoS,
    long = quality-of-service
}

\DeclareAcronym{ism}{
    short = ISM,
    long = {Industrial, Scientific, and Medical}
}

\DeclareAcronym{gsm}{
    short = GSM,
    long = Global System for Mobile communications
}

\DeclareAcronym{wcdma}{
    short = WCDMA,
    long = Wideband Code Division Multiple Access
}

\DeclareAcronym{ble}{
    short = BLE,
    long = Bluetooth Low Energy
}

\DeclareAcronym{if}{
    short = IF,
    long = Instantaneous-Frequency
}

\DeclareAcronym{tf}{
    short = TF,
    long = Time-Frequency
}

\DeclareAcronym{dae}{
    short = DAE,
    long = Deep Auto-Encoder
}

\DeclareAcronym{phoh}{
    short = PHY-OH,
    long = Physical Overhead
}

\begin{document}

\receiveddate{XX Month, XXXX}
\reviseddate{XX Month, XXXX}
\accepteddate{XX Month, XXXX}
\publisheddate{XX Month, XXXX}
\currentdate{29 September, 2026}
\doiinfo{OJCOMS.2026.XXXXXX}

\title{Entropy-Based Wireless Technology Recognition}

\author{
Mohammad Alhellani\IEEEauthorrefmark{1},
Eli De Poorter\IEEEauthorrefmark{1},
Mamoun Guenach\IEEEauthorrefmark{2},
and Adnan Shahid\IEEEauthorrefmark{1}
}

\affil{Internet Technology and Data Science Laboratory (IDLab), Ghent University - IMEC, Ghent 9052, Belgium}

\affil{Interuniversity Microelectronics Centre (IMEC), Leuven 3000, Belgium}

\corresp{CORRESPONDING AUTHOR: Mohammad Alhellani (e-mail: mohammad.alhellani@ugent.be).}
\authornote{This work was supported by the Horizon Europe program under the MCSA Staff Exchanges 2021 grant agreement 101086218 (EVOLVE project); the Horizon-JU-SNS-2023 Research and Innovation Program under Grant Agreement No. 101139194 (6G-XCEL project); and imec.icon project RAPIDNESS, which is co-financed by imec and Flanders Innovation and Entrepreneurship under project nr HBC.2024.0772.}

\begin{abstract}
    \ac{wtr} enables the identification of wireless technologies from received radio signals and is an important capability for spectrum monitoring and management systems. Existing approaches either rely on handcrafted signal features with limited robustness or on computationally intensive \ac{dl} models operating on high-dimensional \ac{iq} samples or spectrograms. In this paper, we investigate a different hypothesis: wireless technologies can be distinguished through the temporal evolution of their local time-frequency complexity. To capture this behavior, we propose a lightweight \ac{wtr} framework based on two-dimensional dispersion entropy. Entropy values are extracted from successive spectrogram regions to characterize the evolution of local spectrogram complexity throughout a transmission and are subsequently summarized into a compact feature representation for classification using an \ac{mlp}. The framework is evaluated on seven wireless technologies, namely IEEE 802.11ah, IEEE 802.15.4 SUN-FSK, IEEE 802.15.4 SUN-OFDM, LoRa, IEEE 802.11n, LTE, and 5G NR, under \acp{snr} ranging from $-20$~dB to $40$~dB. Experimental results show that the proposed approach consistently outperforms \ac{iq}-based and \ac{stft}-based \ac{cnn} baselines, particularly at low and moderate \ac{snr} values. Averaged across all evaluated \ac{snr} levels, it achieves an accuracy of 82.73\%, corresponding to improvements of 23.8\% and 14.6\% over the \ac{iq}-based and \ac{stft}-based \ac{cnn} approaches, respectively. Furthermore, the proposed approach maintains its recognition advantage when data-augmented training is employed while requiring substantially shorter training times. The results demonstrate that the temporal evolution of local spectrogram complexity provides a robust technology-specific fingerprint and an effective, computationally efficient representation for \ac{wtr}.
\end{abstract}


\maketitle

\section{Introduction}

The rapid growth of wireless communication systems has led to increasingly dense and heterogeneous radio environments. Modern wireless deployments support a wide range of applications, including \ac{iot} networks, industrial automation, smart cities, \acp{its}, and broadband communication services. These applications rely on diverse wireless technologies with different physical-layer designs, including operating frequencies, channel bandwidths, and latency requirements. Consequently, multiple wireless technologies are often deployed within the same geographical area and may operate in partially or completely shared spectrum bands. For example, the sub-GHz \ac{ism} band accommodates technologies such as LoRa, IEEE 802.11ah, IEEE 802.15.4 SUN-FSK, and IEEE 802.15.4 SUN-OFDM. Similarly, Wi-Fi and 5G NR-U may coexist in portions of the $5$ GHz and $6$ GHz bands operating in unlicensed spectrum. As wireless deployments continue to expand, the coexistence of heterogeneous technologies in these spectrum bands becomes increasingly common \cite{doi:10.1177/1550147717703966}.

Conventional spectrum-sensing techniques, such as energy detection \cite{7047887}, can determine whether a channel is occupied but cannot identify the transmitting technology. This distinction is important because wireless technologies employ different transmission characteristics and spectrum-access mechanisms, resulting in different interference and coexistence behaviors. Technology recognition therefore provides information beyond channel-occupancy detection and can support spectrum-planning and interference-management decisions. \ac{wtr} addresses this need by identifying the wireless technology associated with an observed waveform and has consequently become an important capability for intelligent spectrum-monitoring and spectrum-management systems.

Early \ac{wtr} systems relied on expert-defined rules or handcrafted signal features, such as \ac{rssi}, bandwidth, periodicity, cyclostationary behavior, spectral features, or temporal activity patterns, to distinguish wireless technologies \cite{article,8844719}, \cite{8570750}. These approaches offer compact representations making them attractive for resource-constrained platforms. However, their performance depends on the careful design of features that capture technology-specific characteristics, which may limit their robustness and scalability in heterogeneous wireless environments.

More recently, \ac{wtr} research has increasingly adopted \ac{dl} models that learn representations directly from raw \ac{iq} samples, \ac{fft}, \ac{stft}, or other signal representations \cite{8292183,8824856}, \cite{9348566,8935690}. \ac{dl}-based approaches have demonstrated excellent recognition performance, particularly with \acp{cnn}, transformer architectures, and \acp{wfm} \cite{11264506}. However, these gains are often accompanied by increased model complexity, large numbers of trainable parameters, and the need for extensive labeled training datasets, resulting in higher computational, memory, and deployment costs.

The complementary strengths and limitations of feature-based and \ac{dl}-based \ac{wtr} methods motivate the development of lightweight frameworks that combine compact signal representations with strong discriminative capability. Rather than relying on large neural networks to learn features directly from high-dimensional signal representations, an alternative approach is to design informative descriptors that capture technology-dependent waveform characteristics while maintaining low computational complexity. Such representations reduce the dimensionality of the classifier input, decrease the number of trainable parameters, and improve the suitability of \ac{wtr} for resource-constrained spectrum-monitoring platforms.

We hypothesize that wireless technologies exhibit distinct temporal evolutions of local time-frequency complexity that can serve as technology-specific fingerprints. To capture this behavior, we propose a lightweight entropy-based \ac{wtr} framework that computes two-dimensional dispersion entropy over localized spectrogram regions and summarizes the resulting complexity evolution using a compact statistical representation. The resulting features preserve technology-dependent waveform characteristics while enabling efficient and robust classification.

The main contributions of this work are summarized as follows:

\begin{itemize} 
\item We demonstrate that the temporal evolution of local spectrogram complexity provides a robust technology-specific fingerprint for wireless technology recognition. 
\item We propose a lightweight entropy-based \ac{wtr} framework based on two-dimensional dispersion entropy and compact statistical feature extraction. 
\item We validate the proposed approach on seven wireless technologies and show improved low-\ac{snr} performance and lower training complexity compared with \ac{iq}-based and \ac{stft}-based \ac{cnn} baselines.
\end{itemize}

The remainder of this paper is organized as follows. Section~II reviews related work. Section~III defines the \ac{wtr} system model, problem formulation, and dataset description. Section~IV presents the methodology, including packet detection, time-frequency processing, entropy-based feature extraction, and classification. Section~V presents and discusses the obtained results. Finally, Section~VI concludes the paper.

\begin{table*}[t] 
\caption{Comparison of Representative 
\ac{wtr} Methods} 
\label{tab:related_work} 
\centering 
\renewcommand{\arraystretch}{1.15} \begin{tabular}{c c c c c} 
\toprule 
\textbf{Ref.} & 
\textbf{Category} & 
\textbf{\# Tech.} & 
\textbf{Features} & 
\textbf{Model} \\ 
\midrule

\cite{8844719} & Feature-Based & 3 & \begin{tabular}[c]{@{}c@{}} 
SCF\\ 
\ac{fft}\\ 
ACF\\ 
PSD 
\end{tabular} 
& \ac{svm} \\ 
\midrule 

\cite{8570750} & Feature-Based & 4 & Spectral + time and power + CCA-based & Classification Tree / Multiclass \ac{svm} \\ 
\midrule 

\cite{s21134286} & Feature-Based & 3 & \ac{if} + Singular Value Decomposition & Random Forest Classifier \\ 
\midrule 

\cite{10049542} & Feature-Based & 7 & DMD & Template Matching \\ 
\midrule 

\cite{8292183} & \ac{dl} & 3 & Internally learned from the spectrogram & \ac{cnn} \\ 
\midrule 

\cite{8824856} & \ac{dl} & 8 & Internally learned from \ac{iq} and \ac{fft} & \ac{cnn} \\ 
\midrule 

\cite{9348566} & \ac{dl} & 6 & Internally learned from \ac{iq} and \ac{fft} & \ac{cnn} \\ 
\midrule 

\cite{8935690} & \ac{dl} & 16 & Internally learned from \ac{iq} & Deep Auto-Encoder + Deep Neural Network \\ 
\midrule 

\cite{11264506} & \ac{dl} & 8 & Internally learned from \ac{iq} and CIR & Transformer \\ 
\midrule 

Proposed & Feature-Based & 7 & Statistical Features of Entropy of Spectrogram & MLP \\ 
\bottomrule 
\end{tabular} 
\end{table*}

\section{Related Work}

\ac{wtr} has been investigated using different approaches. Existing methods primarily differ in how signal characteristics are represented and how the resulting representations are classified.

Early feature-based approaches relied on manually designed signal descriptors combined with conventional \ac{ml} classifiers. For example, \cite{8844719} jointly exploited Spectral Correlation Function (SCF), \ac{fft}, Auto-Correlation Function (ACF), and Power Spectral Density (PSD) features with \acp{svm} to distinguish \ac{gsm}, \ac{wcdma}, and \ac{lte}. In \cite{8570750}, envelope and model-assisted spectral features were combined with conventional classifiers for recognition of IEEE 802.11b/g/n, 802.15.4, 802.15.1, and \ac{ble}, achieving sub-ms inference latency on commercial \ac{iot} hardware. The work in \cite{s21134286} compressed \ac{if} features obtained from time-frequency analysis using Singular Value Decomposition before classification with a Random Forest, while \cite{10049542} employed Dynamic Mode Decomposition (DMD) to capture periodic signal characteristics for recognition of seven wireless technologies.

More recent approaches employ \ac{dl} to learn discriminative representations directly from raw \ac{iq} samples or transformed signal representations. \acp{cnn} have been applied to spectrograms, \ac{iq} samples, and \ac{fft} representations for recognition of multiple wireless technologies \cite{8292183,8824856}, \cite{9348566}. Semi-supervised learning has further reduced the dependence on labeled data while preserving high recognition accuracy across multiple wireless technologies \cite{8935690}. More recently, \acp{wfm} have enabled a unified approach to wireless tasks, including \ac{wtr}, with improved generalization to unseen technologies and propagation environments \cite{11264506}. Table~\ref{tab:related_work} summarizes representative \ac{wtr} methods discussed in this section, comparing their learning paradigm, feature representation, and classification model.

Despite their strong recognition performance, \ac{dl}-based methods generally require more computational resources, memory, and trainable parameters than feature-based approaches. This can limit their suitability for lightweight spectrum-monitoring platforms. Feature-based methods therefore remain attractive when computational efficiency and limited training requirements are important.

The existing feature-based approaches primarily characterize individual or aggregated properties of the received waveform, such as spectral content, bandwidth, temporal activity, periodicity, or cyclostationary structure. However, these descriptors do not explicitly characterize how signal complexity is distributed and evolves across the joint time-frequency domain. Wireless technologies differ in how energy is organized, concentrated, and dispersed over time and frequency as a consequence of their physical-layer waveforms. The technology-dependent structure of these waveforms determines how signal energy is distributed across the time-frequency plane, resulting in distinct local levels of regularity and irregularity in spectrograms. Measuring this local complexity through entropy therefore provides a complementary characterization of the waveform: rather than describing \emph{what spectral or temporal components} are present, it quantifies \emph{how their local complexity} evolves over time. This distinction motivates the use of local time-frequency entropy as a potential technology-specific fingerprint.

\section{System Model, Problem Formulation and Dataset Description}

This section presents the considered \ac{wtr} system model, formulates the corresponding classification problem, and describes the dataset used for evaluation. The system model and deployment scenarios are first introduced, followed by the received signal model and mathematical formulation of the \ac{wtr} problem. Finally, the considered wireless technologies and the main characteristics of the experimental dataset are described.

\subsection{System Model}

In this work, we consider a \ac{wtr} system that identifies heterogeneous wireless technologies by extracting discriminative features from received \ac{iq} observations. The considered technologies are represented by the class
\[
\begin{aligned}
\mathcal{C}=
\{&
\text{LoRa},\,
\text{IEEE~802.11ah},\
\text{IEEE~802.15.4 SUN-FSK},\\
&
\text{IEEE~802.15.4g SUN-OFDM},\,
\text{IEEE~802.11n},\\
&
\text{LTE},\,
\text{5G NR}
\},
\end{aligned}
\]

The considered technologies operate across different frequency ranges
and may coexist in shared or overlapping spectral regions. For
example, LoRa and IEEE~802.15.4 SUN-FSK considered in this work are
both centered at 868.1~MHz, with bandwidths of 125~kHz and 200~kHz,
respectively, resulting in direct spectral overlap. In the GHz range,
coexistence between 5G NR-U and Wi-Fi systems has also been extensively
studied, particularly in the $5$~GHz and $6$~GHz bands, where
simultaneous operation can result in channel contention and
interference \cite{KOSEKSZOTT2021108188,9488780}. These examples
illustrate that heterogeneous wireless technologies may operate in
spectrally shared environments, motivating the need to identify the
active technology before appropriate coexistence or spectrum-management decisions can be made.

The objective of the \ac{wtr} system is to identify the active
wireless technology directly from the received \ac{iq} samples,
without demodulating or decoding the transmitted signal and without
requiring protocol-specific receivers. The \ac{iq} samples are
processed to extract a compact entropy-based representation for
\ac{wtr}. The identified technology can subsequently be used to
support spectrum-aware coexistence and resource-allocation decisions.

\subsection{Problem Formulation}
\label{subsec:ProblemFormulation}

\ac{wtr} is formulated as a supervised multi-class classification
problem, where each class corresponds to one of the considered
wireless technologies. The objective is to determine the wireless technology
$c\in\mathcal{C}$ associated with a received signal.

The received complex baseband signal is modeled as

\begin{equation}
r[n]
=
s[n]
e^{j2\pi \frac{\Delta f}{F_s}n}
+
w[n]
=
I[n]
+
jQ[n],
\label{eq:signal_model}
\end{equation}
where $r[n]$ denotes the received complex baseband signal,
$s[n]$ is the transmitted signal, $\Delta f$ is the \ac{cfo}, $F_s$ is the sampling frequency,
and $w[n]\sim\mathcal{N}(0,\sigma_w^2)$ denotes \ac{awgn}.

Rather than performing classification directly on the received
\ac{iq} samples, the proposed framework maps the received
signal to a compact feature vector through a feature-extraction function,

\begin{equation}
\Phi:\;
r[n]
\rightarrow
\mathbf{x}
\in
\mathbb{R}^{d},
\end{equation}
where $\mathbf{x}$ denotes the extracted feature vector and
$d$ is its dimension.

The feature vector is subsequently classified by a lightweight
\ac{mlp},

\begin{equation}
g_{\boldsymbol{\theta}}:
\mathbb{R}^{d}
\rightarrow
\mathcal{C},
\end{equation}
where $\boldsymbol{\theta}$ denotes the trainable parameters
of the classifier.

The classifier is trained in a supervised manner using the
dataset

\begin{equation}
\mathcal{D}
=
\left\{
(\mathbf{x}_i,c_i)
\right\}_{i=1}^{N},
\end{equation}
where $N$ denotes the number of training samples,
$\mathbf{x}_i$ is the extracted feature vector of the $i$th
received signal, and $c_i\in\mathcal{C}$ is its corresponding
ground-truth technology label.

The optimal classifier parameters are obtained by minimizing
the empirical cross-entropy loss,

\begin{equation}
\boldsymbol{\theta}^{*}
=
\arg\min_{\boldsymbol{\theta}}
\frac{1}{N}
\sum_{i=1}^{N}
\mathcal{L}
\!\left(
c_i,
g_{\boldsymbol{\theta}}(\mathbf{x}_i)
\right),
\end{equation}
where $\mathcal{L}(\cdot,\cdot)$ denotes the cross-entropy
loss function.

\subsection{Dataset Description}
\label{subsec:DatasetDescription}

The experimental dataset comprises seven wireless technologies, with each technology treated as a separate classification class. The considered technologies are IEEE 802.11ah, IEEE 802.15.4 SUN-FSK, IEEE 802.15.4 SUN-OFDM, LoRa, LTE, 5G NR, and IEEE 802.11n. The IEEE 802.11ah, IEEE 802.15.4 SUN-FSK, IEEE 802.15.4 SUN-OFDM, LoRa, and LTE datasets are obtained from the publicly available IDLab wireless signal repository~\cite{9348566}, whereas the IEEE 802.11n dataset is obtained from~\cite{72qq-z464-23}. The 5G NR dataset is captured using the IDLab cutting-edge 5G/6G testbed \footnote{\url{https://idlab.ugent.be/resources/cutting-edge-5g-6g-research-testbed}}. Table~\ref{tab:dataset} summarizes the main characteristics of the considered wireless technologies, including their center frequency, bandwidth, and sampling rate.

To evaluate the recognition performance under different channel conditions, \ac{awgn} is added to the original recordings to generate signals with \ac{snr} values ranging from $-20$ dB to $40$ dB in $5$-dB increments. The original datasets consist of continuous \ac{iq} recordings containing wireless signal transmissions. Packet boundaries are identified using the magnitude-threshold packet detection algorithm described in Subsection~\ref{subsec:PacketDetection}. The detected packets are then divided into training and testing sets based on their order in the original recordings. Specifically, the odd-numbered detected packets are assigned to the training set, while the even-numbered packets are assigned to the testing set.

\begin{table}[t]
\caption{Characteristics of the Wireless Technologies Included in the Experimental Dataset}
\label{tab:dataset}
\centering
\footnotesize
\renewcommand{\arraystretch}{1.15}
\begin{tabular}{lccc}
\hline
\textbf{Technology} &
\textbf{$f_{\mathrm{center}}$ (MHz)} &
\textbf{Bandwidth (MHz)} &
\textbf{$F_s$ (MHz)} \\
\hline

\multirow{4}{*}{IEEE 802.11ah}
& 863.5  & 1  & \multirow{4}{*}{2.048 } \\
& 864.0  & 2  & \\
& 864.5  & 1  & \\
& 866.0  & 2  & \\
\hline

\noalign{\vskip 1.5mm}
\shortstack[c]{IEEE 802.15.4\\SUN-FSK}
& 868.1  & 0.2 & 2.048  \\
\hline

\multirow{4}{*}{\shortstack[c]{IEEE 802.15.4\\SUN-OFDM}}
& 863.625  & 1.2  & \multirow{4}{*}{2.048 } \\
& 863.425  & 0.8  & \\
& 863.225  & 0.4  & \\
& 863.125  & 0.2  & \\
\hline

LoRa
& 868.1  & 0.125  & 2.048  \\
\hline

LTE
& 5240  & 1  & 1  \\
\hline

5G NR
& 2647.5  & 10  & 10  \\
\hline

IEEE 802.11n
& 5885 & 20  & 20  \\
\hline

\end{tabular}
\end{table}

\section{Proposed Wireless Technology Recognition Method}
\label{sec:framework}

\begin{figure*}[!t]
    \centering
    \includegraphics[width=\textwidth]{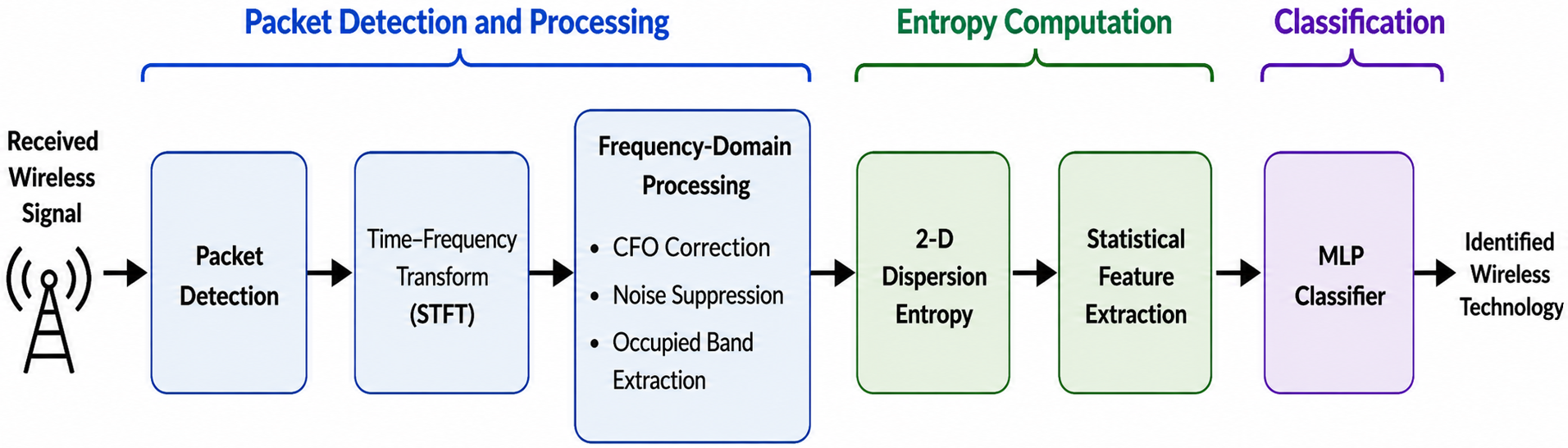}
    \caption{Overview of the proposed entropy-based \ac{wtr} framework}
    \label{fig:pipeline}
\end{figure*}

This section presents the proposed \ac{wtr} framework. Fig.~\ref{fig:pipeline} provides an overview of the complete processing pipeline, which transforms the received \ac{iq} stream into wireless technology predictions through packet detection, time-frequency transformation, frequency-domain processing, entropy computation, feature extraction, and supervised classification. The following subsections describe each processing stage in detail.

\subsection{Packet Detection and Observation Window}
\label{subsec:PacketDetection}

Following the baseband signal model in \eqref{eq:signal_model}, a magnitude-threshold packet detection is applied to the continuous \ac{iq} stream to extract the packets. This step ensures that the subsequent time-frequency and entropy-based analysis is performed on the detected packets.

Let $a[n] = |r[n]|$ denote the magnitude of the received complex baseband signal. A sample is classified as active when its magnitude exceeds a predefined detection threshold $\gamma$. The threshold is selected to be equal to the measured noise floor in order to distinguish signal samples from background noise. The total noise power observed by the receiver can be expressed as $P_{\mathrm{noise}} = P_{\mathrm{env}} + P_{\mathrm{rx}}$ where $P_{\mathrm{env}}$ represents the environmental noise power received by the antenna and $P_{\mathrm{rx}}$ denotes the receiver
noise power. The latter can be expressed as $P_{\mathrm{rx}} = kTBF$ where $k$ is Boltzmann's constant, $T$ is the absolute temperature, $B$ is the receiver bandwidth, and $F$ is the
receiver noise factor.

The environmental noise power $P_{\mathrm{env}}$ includes
background radio-frequency emissions, man-made electromagnetic
noise, and other noise sources present in the
surrounding environment. Since this contribution varies with
the operating conditions, the overall
noise level can't generally be determined accurately from the
receiver parameters alone. Therefore, the threshold $\gamma$ is estimated empirically from idle-channel observations by measuring the \ac{iq} magnitude during intervals in which no signal is transmitted. This approach automatically captures the combined effects of receiver noise and environmental noise.

Consecutive active samples are grouped to form a detected packet. Due to noise and modulation-dependent amplitude variations, the received signal may briefly fall below $\gamma$ even while the transmission is ongoing. To avoid fragmenting a single packet into multiple detections, below-threshold intervals shorter than $G$ samples are not considered packet boundaries. Consequently, a packet is declared to have ended only when the signal remains below the detection threshold for at least $G$ consecutive samples. The parameter $G$ was selected empirically to suppress short inactive intervals occurring within a packet while remaining sufficiently small to avoid merging consecutive packets.

The packet detector returns the start and end sample indices of
each detected packet, denoted by $n_s$ and $n_e$, respectively.
The detected packet length is therefore $n_e-n_s+1$ samples.

Processing the entire packet is not always necessary for reliable technology recognition. Many wireless technologies exhibit characteristic structures at the beginning of a transmission. For packet-based technologies, such as LoRa and IEEE~802.11ah, the initial portion typically contains technology-dependent components, including the preamble and synchronization fields. Similarly, frame-based technologies, such as \ac{lte} and \ac{5g}, exhibit recurring frames. Consequently, after a representative portion of the transmission has been observed,
additional samples often provide limited discriminative information while increasing the computational cost.

To exploit this observation, each detected transmission is
restricted to a maximum observation duration $T_w$. If a transmission lasts longer than $T_w$, only the first $T_w$ duration of the detected signal are considered. The
number of retained samples is defined as

\begin{equation}
L =
\min \left( n_e - n_s + 1,\; T_w F_s \right).
\label{eq:L}
\end{equation}

The retained observation is then expressed as

\begin{equation}
\mathbf{r}
=
[r[n_s],\, r[n_s+1],\, \ldots,\, r[n_s+L-1]]^{T}.
\label{eq:packet_vector}
\end{equation}

Since the wireless technology is unknown at the receiver, the
observation window should be sufficiently long to capture a
representative time-frequency characteristic of each considered
technology while avoiding unnecessarily long observation intervals.
For each technology $c\in\mathcal{C}$, let
$T_{\mathrm{rep},c}$ denote the minimum temporal duration required to
capture a representative time-frequency characteristic of technology
$c$. The observation window is therefore selected according to
\begin{equation}
T_w \geq \max_{c\in\mathcal{C}} T_{\mathrm{rep},c}.
\label{eq:observation_window}
\end{equation}

\begin{table}[t]
\centering
\caption{Representative temporal structures used to determine the
observation-window duration.}
\label{tab:observation_window}
\begin{tabular}{l c c}
\hline
\textbf{Technology} & \textbf{Representative structure} &
\textbf{$T_{\mathrm{rep},c}$ (ms)}\\
\hline
IEEE 802.11ah & PHY-OH & $0.560$ \\
\hline
IEEE 802.11n & PHY-OH & $0.052$ \\
\hline
\noalign{\vskip 1.5mm}
\shortstack[c]{IEEE 802.15.4\\SUN-OFDM} & PHY-OH & $1.68$ \\
\hline
LoRa & Complete chirp symbol & $32.768$ \\
\hline
\noalign{\vskip 1.5mm}
\shortstack[c]{IEEE 802.15.4\\SUN-FSK} & Complete octet & $0.160$ \\
\hline
LTE & Complete frame & $10$ \\
\hline
5G NR & Complete frame & $10$ \\
\hline
\end{tabular}
\end{table}

The representative duration is determined according to the
underlying waveform and temporal structure of each technology, as
summarized in Table~\ref{tab:observation_window}. For IEEE~802.11ah,
IEEE~802.11n, and IEEE~802.15.4 SUN-OFDM, the complete \ac{phoh} is considered the representative structure, since it contains
characteristic time-frequency patterns associated with the
transmitted waveform. For LoRa, the representative structure is a
complete chirp symbol, whose time-frequency evolution characterizes
the chirp-based waveform. For IEEE~802.15.4 SUN-FSK, the representative structure is a
complete octet, corresponding to eight transmitted bits whose symbols produce a sequence of frequency transitions associated with the 2-FSK or 4-FSK modulation. For LTE
and 5G NR, one complete frame is considered sufficient to capture
the characteristic temporal and time-frequency structure of the
transmission.

The duration reported for each technology in Table~\ref{tab:observation_window} represents the
maximum possible duration of the corresponding representative
structure over the configurations specified for that technology. For example, the duration of
a LoRa chirp symbol depends on the spreading factor and bandwidth;
therefore, the value reported for LoRa corresponds to the longest
symbol duration among the considered parameter configurations. The
same principle is applied to technologies whose representative
structures depend on configurable waveform parameters. The representative durations obtained for the considered technologies are reported in Table~\ref{tab:observation_window}.
Based on these requirements, an observation window of
$T_w=33~\mathrm{ms}$ is adopted. This duration provides sufficient
time-frequency evolution to characterize all considered technologies
while maintaining a fixed, technology-agnostic observation duration
for the subsequent time-frequency and entropy-based feature
extraction stages.

If a detected packet is shorter than $T_w=33~\mathrm{ms}$, the
complete packet is used without padding. Thus, the retained
observation in \eqref{eq:packet_vector} provides a practical input
for the subsequent time-frequency and entropy-based feature
extraction stages.

\subsection{Time-Frequency Representation}

The retained packet observation $\mathbf r$ contains both temporal and spectral information. However, spectral representations often provide more discriminative and noise-robust features than raw time-domain waveforms at low \ac{snr}, since they concentrate signal energy in localized frequency regions while broadband noise is more uniformly distributed \cite{ozturk2020rfbasedlowsnrclassificationuavs}. Therefore, each detected packet is transformed into a time-frequency representation using the \ac{stft}.

Let $g[\ell]$ denote an analysis window of length $N_w$, and let $H$ be the
hop size between consecutive \ac{stft} frames. The \ac{stft} of the retained packet
$\mathbf r$ is defined as
\begin{equation}
X[k,m]
=
\sum_{\ell=0}^{N_w-1}
r[\ell+mH] g[\ell]
e^{-j2\pi k\ell/N_w},
\label{eq:stft}
\end{equation}
where $k$ denotes the frequency-bin index and $m$ denotes the time-frame index.
The hop size is given by $H = N_w - N_o$ where $N_o$ is the number of overlapping samples between adjacent analysis windows.

The corresponding time-frequency power representation is obtained as
\begin{equation}
S[k,m] = |X[k,m]|^2.
\label{eq:spectrogram}
\end{equation}

The matrix $\mathbf S$ formed by $S[k,m]$ represents the power distribution over frequency and time. This representation is used as the basis for the subsequent frequency-domain processing and entropy-based feature extraction stages.

The STFT window length was selected based on a sensitivity analysis, where $N_w=128$ samples achieved the same recognition performance as smaller window lengths, while larger windows (e.g. 256 samples) resulted in performance degradation. Thus, a Hann window with $N_w=128$ samples was selected to maintain recognition performance while reducing computational cost. An overlap of $N_o=64$ samples is used between consecutive frames.

\subsection{Frequency-Domain Processing}

The \ac{stft} power matrix obtained in the previous subsection represents the time-frequency distribution of the retained packet observation. However, the received signal may be affected by \ac{cfo} and noise. \ac{cfo} can arise from mismatches between the transmitter and receiver local oscillators and, in mobile scenarios, from Doppler effects. \ac{cfo} correction is particularly important because a sufficiently large frequency offset can shift part of the signal spectrum toward the boundaries of the spectrogram. Consequently, a single occupied band may appear as two disconnected spectral regions, which can distort the subsequent time-frequency analysis and entropy extraction. In addition, thermal noise, receiver noise, and other sources of interference may obscure or distort the time-frequency characteristics used by the proposed entropy representation. Since the entropy extraction relies on these local time-frequency characteristics to distinguish between wireless technologies, noise-dominated components can introduce irrelevant patterns and reduce the discriminative information of the resulting representation.

To reduce the influence of such components, the frequency-domain processing stage, therefore, first compensates the \ac{cfo} and then suppresses the noise-dominated regions while retaining the dominant signal components that contain the time-frequency characteristics relevant to wireless technology recognition.

In the current implementation, a preliminary estimate of \ac{cfo} is obtained from the spectrogram using a power-weighted spectral centroid. Let $f_k$ denote the frequency associated with the $k$th frequency bin. The average spectral power is computed as
\begin{equation}
\bar{S}[k]
=
\frac{1}{M}
\sum_{m=0}^{M-1}
S[k,m],
\label{eq:average_spectral_power}
\end{equation}
where $M$ is the number of \ac{stft} time frames per detected packet. The \ac{cfo} is then estimated as
\begin{equation}
\widehat{\Delta f}
=
\frac{
\sum_k f_k \bar{S}[k]
}{
\sum_k \bar{S}[k]
}.
\label{eq:cfo_estimation}
\end{equation}

The retained packet observation is compensated in the time
domain according to
\begin{equation}
\tilde r[n]
=
r[n]
e^{-j2\pi \frac{\widehat{\Delta f}}{F_s}n}.
\label{eq:cfo_correction}
\end{equation}

It should be noted that the objective of this stage is not to obtain a highly accurate \ac{cfo} estimate. Instead, the estimated offset provides a coarse recentring of the occupied spectrum prior to occupied-band extraction. Even when the estimated offset is not fully accurate, the subsequent adaptive occupied-band extraction stage remains capable of isolating the dominant spectral occupancy of the received waveform.
The \ac{stft} is then recomputed from $\tilde r[n]$, and the corresponding power spectrogram is again denoted by $\mathbf S$ for notational simplicity.

After \ac{cfo} compensation, low-power time-frequency coefficients are suppressed
using a relative threshold $\rho$. The thresholded spectrogram is defined as
\begin{equation}
S_{\mathrm{th}}[k,m]
=
S[k,m]
\mathbb{I}
\left(
S[k,m]
\geq
\rho
\max_{k,m} S[k,m]
\right),
\label{eq:spectrogram_thresholding}
\end{equation}
where $\rho$ is the spectrogram threshold ratio and $\mathbb{I}(\cdot)$ is the
indicator function. The threshold ratio was selected through a sensitivity analysis, where $\rho$ was varied from $0$ to $0.4$ in increments of $0.05$. Among the evaluated values, $\rho=0.15$ provided the best recognition performance while effectively suppressing low-power, noise-dominated components. Therefore, $\rho=0.15$ was adopted throughout the experiments.

The occupied frequency region is then extracted adaptively from the thresholded spectrogram. This step restricts the subsequent entropy analysis to the frequency region containing the dominant transmission energy, thereby reducing the influence of inactive or noise-dominated frequency bins. Similar to normalization, this operation provides a common frequency-domain reference across transmissions with different occupied bandwidths. First, the average power of each frequency bin is computed as
\begin{equation}
\bar{S}_{\mathrm{th}}[k]
=
\frac{1}{M}
\sum_{m=0}^{M-1}
S_{\mathrm{th}}[k,m].
\label{eq:frequency_power_profile}
\end{equation}

The set of active frequency bins is defined as
\begin{equation}
\mathcal K
=
\left\{
k:
\bar{S}_{\mathrm{th}}[k]
\geq
\eta
\max_n \bar{S}_{\mathrm{th}}[n]
\right\},
\label{eq:active_frequency_bins}
\end{equation}
where $\eta$ is the occupied-band threshold ratio. The value of $\eta$ was selected through a sensitivity analysis in which $\eta$ was varied from $0$ to $0.5$ in increments of $0.05$. A value of $\eta=0.35$ was selected based on the resulting recognition performance and was used throughout the experiments.

Let $k_{\min}=\min \mathcal K,\qquad k_{\max}=\max \mathcal K$. The occupied-band spectrogram is then obtained as
\begin{equation}
\mathbf S_c
=
\mathbf S_{\mathrm{th}}
[
k_{\min}:k_{\max},
:
],
\label{eq:cropped_spectrogram}
\end{equation}
where
\begin{equation}
\mathbf S_c \in \mathbb{R}_{+}^{K_c \times M},
\end{equation}
and $K_c$ is the resulting number of retained frequency bins,

\begin{equation}
K_c = k_{\max} - k_{\min} + 1.
\end{equation}

The resulting matrix $\mathbf S_c$ contains the dominant time-frequency
content of the detected packet within its occupied bandwidth. By removing
inactive spectral regions, this representation reduces the influence of
out-of-band noise and focuses the subsequent entropy computation on the part of
the spectrogram that carries the technology-dependent waveform structure.

\begin{algorithm}[t]
\caption{Entropy Sequence Generation Using 2D Dispersion Entropy}
\label{alg:2dde}
\begin{algorithmic}[1]

\Require Cropped spectrogram $\mathbf{S}_c$, patch length $M_p$, number of classes $C$, embedding dimensions $(m_h,m_w)$
\Ensure Entropy sequence $\mathbf{h}$

\State Compute
$S_{\min}=\min(\mathbf{S}_c)$ and
$S_{\max}=\max(\mathbf{S}_c)$

\State Divide $\mathbf{S}_c$ into non-overlapping time patches
$\{\mathbf{A}_q\}_{q=1}^{Q}$

\For{each patch $\mathbf{A}_q$}

    \State Normalize the spectrogram amplitudes using
    $S_{\min}$ and $S_{\max}$

    \State Quantize the normalized amplitudes into
    $C$ dispersion classes to obtain $\mathbf{Z}_q$

    \State Construct the dispersion patterns by sliding an
    $m_h \times m_w$ embedding window over $\mathbf{Z}_q$

    \State Estimate the probability of each
    dispersion pattern

    \State Compute the entropy $h_q$

\EndFor

\State Form the entropy sequence
$\mathbf{h}=[h_1,h_2,\ldots,h_Q]^T$

\Return $\mathbf{h}$

\end{algorithmic}
\end{algorithm}

\subsection{Two-Dimensional Dispersion Entropy}

Instead of using the spectrogram values directly, the proposed framework characterizes the local complexity of $\mathbf{S}_c$ using two-dimensional dispersion entropy. This representation is particularly suitable for \ac{wtr} because different wireless technologies exhibit distinct local regularity patterns in their spectrograms. Consequently, chirp-based, FSK-based, and OFDM-based waveforms generate different temporal evolutions of spectrogram complexity. By explicitly quantifying these complexity patterns, the proposed entropy representation extracts compact technology-specific fingerprints that can be used for robust wireless technology recognition.

To capture the temporal evolution of the time-frequency complexity, $\mathbf S_c$ is divided along the time axis into non-overlapping patches of length $M_p$ \ac{stft} frames. Computing the entropy over the entire spectrogram would produce a single entropy value and therefore discard information about how the local complexity evolves over time. In contrast, computing the entropy separately for each patch produces a sequence of entropy values that characterizes the temporal evolution of the time-frequency complexity, analogous to the use of \ac{stft} to preserve temporal information that would be lost when applying \ac{fft} to the entire signal.
The $q$th time-frequency patch is denoted by
\begin{equation}
\mathbf A_q =
\mathbf S_c[:,\, (q-1)M_p : qM_p-1],
\label{eq:entropy_patch}
\end{equation}
where
\[
\mathbf A_q \in \mathbb{R}_{+}^{K_c \times M_p}.
\]

A smaller patch length provides finer temporal resolution of the entropy, allowing more detailed characterization of the temporal evolution of the time-frequency randomness. It also reduces the computational complexity of the entropy calculation by decreasing the number of embedding vectors processed within each patch. Based on simulation results, $M_p=5$ provides the best trade-off between the temporal resolution of the entropy sequence and computational complexity. It is also worth noting that this patch length should accommodate the temporal embedding dimension $m_w$.

The entropy value associated with each patch, representing the local dispersion complexity of the occupied-band spectrogram over a short time interval, is computed using the 2D-Dispersion Entropy method proposed in \cite{AZAMI2019178}. The overall procedure for generating the entropy sequence is summarized in Algorithm~\ref{alg:2dde}, while the mathematical formulation of each step is presented below. 

The spectrogram amplitudes are mapped into
a finite number of discrete classes. Let $C$ denote the number of dispersion
classes. The minimum and maximum values of the cropped spectrogram are defined
as
\begin{equation}
S_{\min} = \min_{k,m} S_c[k,m],
\qquad
S_{\max} = \max_{k,m} S_c[k,m].
\label{eq:spectrogram_min_max}
\end{equation}

Each coefficient in the patch $\mathbf A_q$ is normalized with respect to the
dynamic range of the cropped spectrogram and mapped to one of $C$ discrete
dispersion classes as
\begin{equation}
Z_q[k,m]
=
\left\lfloor
C
\frac{
A_q[k,m]-S_{\min}
}{
S_{\max}-S_{\min}+\epsilon
}
\right\rfloor,
\label{eq:dispersion_quantization}
\end{equation}
where $C$ is the number of dispersion classes and $\epsilon$ is a small
positive constant used to avoid division by zero. Since $\epsilon>0$, the
largest spectrogram coefficient is mapped to a value slightly smaller than
$C$, and therefore the floor operation assigns it to the last valid class
$C-1$. Hence, the quantized coefficients satisfy
\[
Z_q[k,m]\in\{0,1,\ldots,C-1\}.
\label{eq:dispersion_class_set}
\]

In this work, the number of dispersion classes is fixed to $C=6$, following the recommendation in \cite{e23101303}, where this value was identified as an appropriate choice for dispersion entropy.

The two-dimensional dispersion patterns are then formed by sliding a local
embedding window over $\mathbf Z_q$. Let $m_h$ and $m_w$ denote the embedding
dimensions along the frequency and time axes, respectively. For a given
position $(u,v)$, the local quantized block is
\begin{equation}
\mathbf B_q^{(u,v)}
=
\mathbf Z_q[u:u+m_h-1,\; v:v+m_w-1],
\label{eq:local_dispersion_block}
\end{equation}
where
\[
u=0,\ldots,K_c-m_h,
\qquad
v=0,\ldots,M_p-m_w.
\]

The corresponding dispersion pattern is obtained by vectorizing this block:
\begin{equation}
\boldsymbol{\pi}_q^{(u,v)}
=
\operatorname{vec}\left(
\mathbf B_q^{(u,v)}
\right),
\label{eq:dispersion_pattern}
\end{equation}
where $\operatorname{vec}(\cdot)$ denotes the vectorization operator that reshapes a matrix into a vector.

Each pattern $\boldsymbol{\pi}_q^{(u,v)}$ is therefore a vector of length
$m_hm_w$, whose entries belong to the set $\{0,1,\ldots,C-1\}$.

Let $\mathcal P$ denote the set of all possible dispersion patterns. The total number of
possible patterns is

\begin{equation}
|\mathcal{P}| = C^{m_hm_w}.
\end{equation}

For the $q$th patch, the number of observed dispersion
patterns is

\begin{equation}
N_q
=
(K_c-m_h+1)(M_p-m_w+1).
\end{equation}

The probability of observing a pattern
$\pi \in \mathcal{P}$ in the $q$th patch is estimated empirically
as

\begin{equation}
p_q(\pi)
=
\frac{1}{N_q}
\sum_{u=0}^{K_c-m_h}
\sum_{v=0}^{M_p-m_w}
\mathbb{I}\left(
\pi_q^{(u,v)}=\pi
\right).
\end{equation}

The two-dimensional dispersion entropy of the $q$th patch is then computed as
\begin{equation}
h_q
=
-
\sum_{\boldsymbol{\pi}\in\mathcal P}
p_q(\boldsymbol{\pi})
\log
\left(
p_q(\boldsymbol{\pi})+\epsilon
\right).
\label{eq:2d_dispersion_entropy}
\end{equation}

A low value of $h_q$ indicates that only a small number of dispersion patterns dominate the local time-frequency patch, corresponding to a more regular structure. In contrast, a high value of $h_q$ indicates a more diverse distribution of local dispersion patterns, corresponding to higher local time-frequency complexity.

\begin{figure*}[!ht]
    \centering

    \begin{tabular}{@{}c@{\hspace{2mm}}c@{\hspace{3mm}}c@{\hspace{3mm}}c@{}}

    &
    \makebox[0.27\textwidth][c]{\large\textbf{LoRa}}
    &
    \makebox[0.27\textwidth][c]{\large\textbf{5G NR}}
    &
    \makebox[0.27\textwidth][c]{\large\textbf{IEEE 802.11ah}}
    \\[-1mm]

    \raisebox{8mm}{\rotatebox{90}{\textbf{Frequency (MHz)}}}
    &
    \shortstack[c]{
        \includegraphics[
            width=0.29\textwidth,
            height=0.20\textwidth
        ]{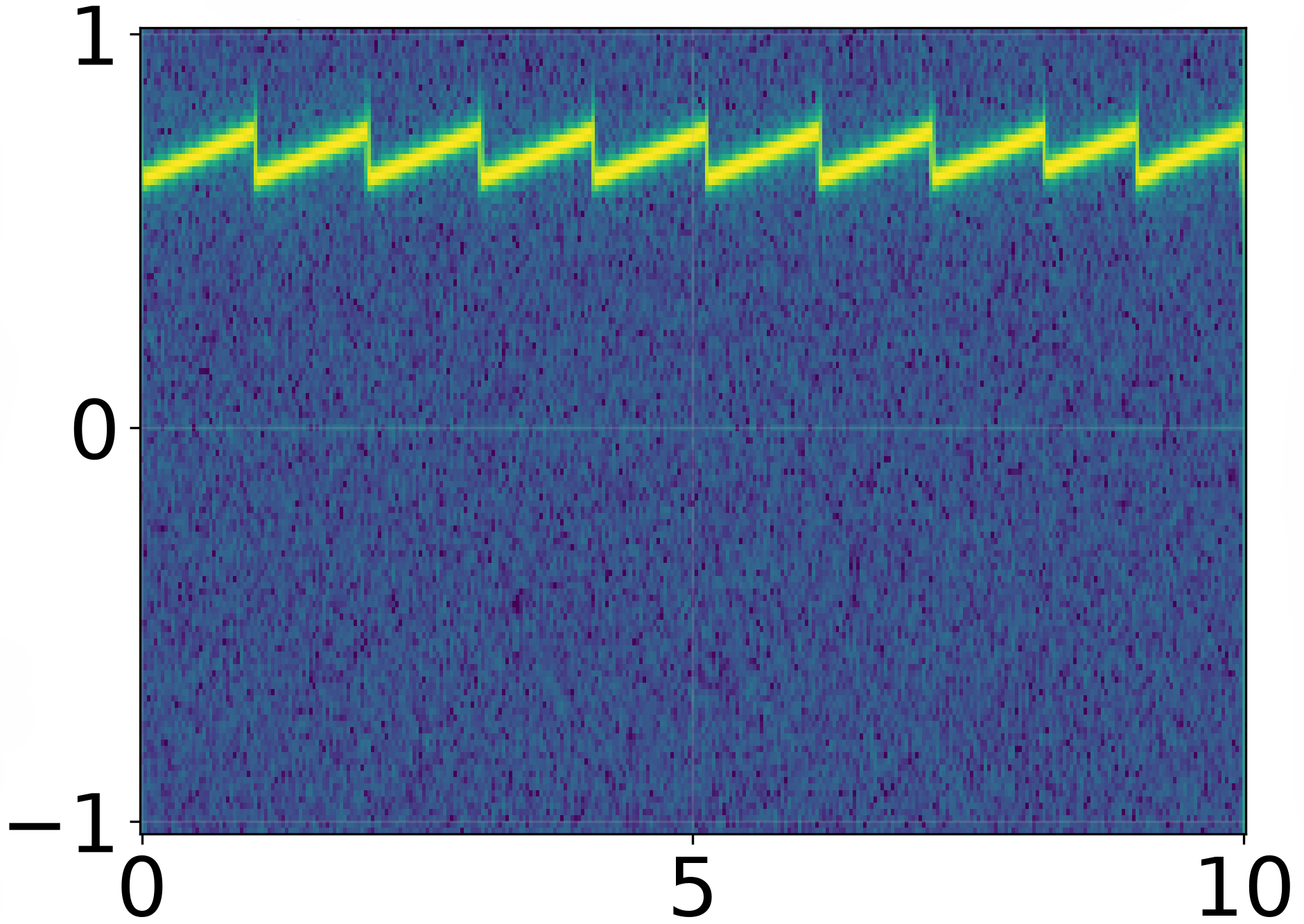}\\[-1mm]
        \normalsize Time (ms)
    }
    &
    \shortstack[c]{
        \includegraphics[
            width=0.30\textwidth,
            height=0.20\textwidth
        ]{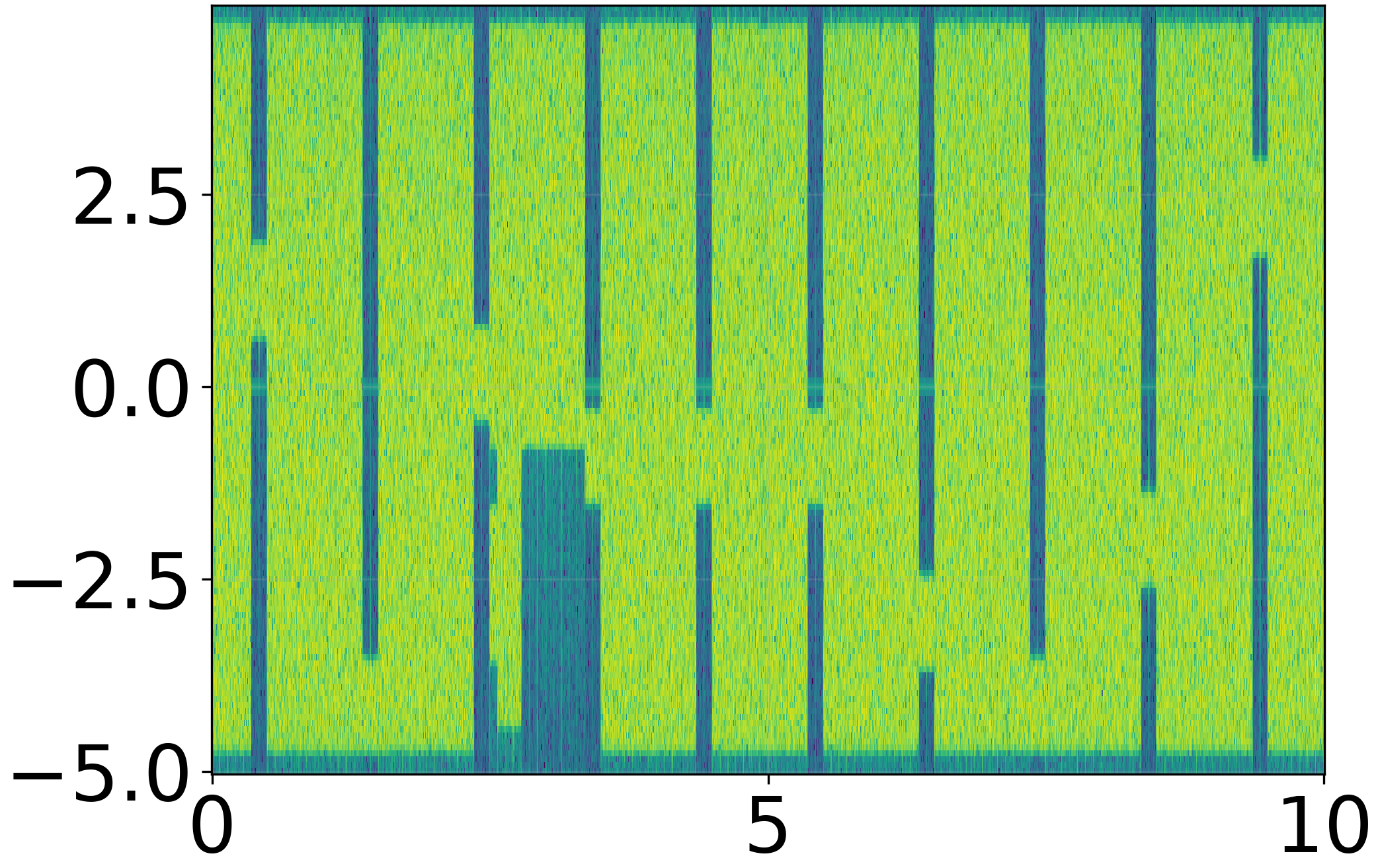}\\[-2mm]
        \normalsize Time (ms)
    }
    &
    \shortstack[c]{
        \includegraphics[
            width=0.34\textwidth,
            height=0.20\textwidth
        ]{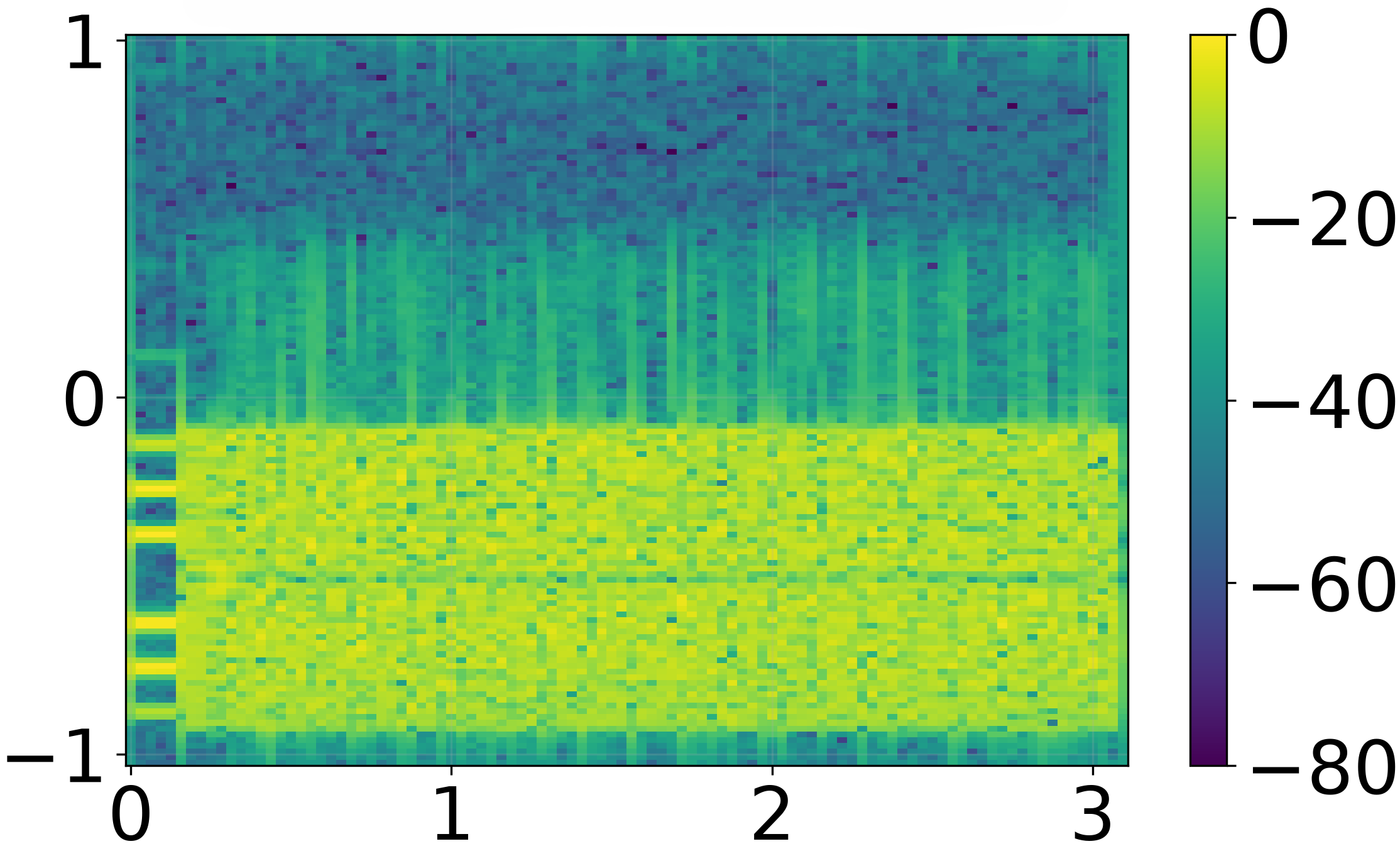}\\[-1mm]
        \normalsize Time (ms)
    }
    \\[4mm]

    \raisebox{14mm}{\rotatebox{90}{\textbf{Entropy}}}
    &
    \shortstack[c]{
        \hspace{2mm}
        \includegraphics[
            width=0.27\textwidth,
            height=0.20\textwidth
        ]{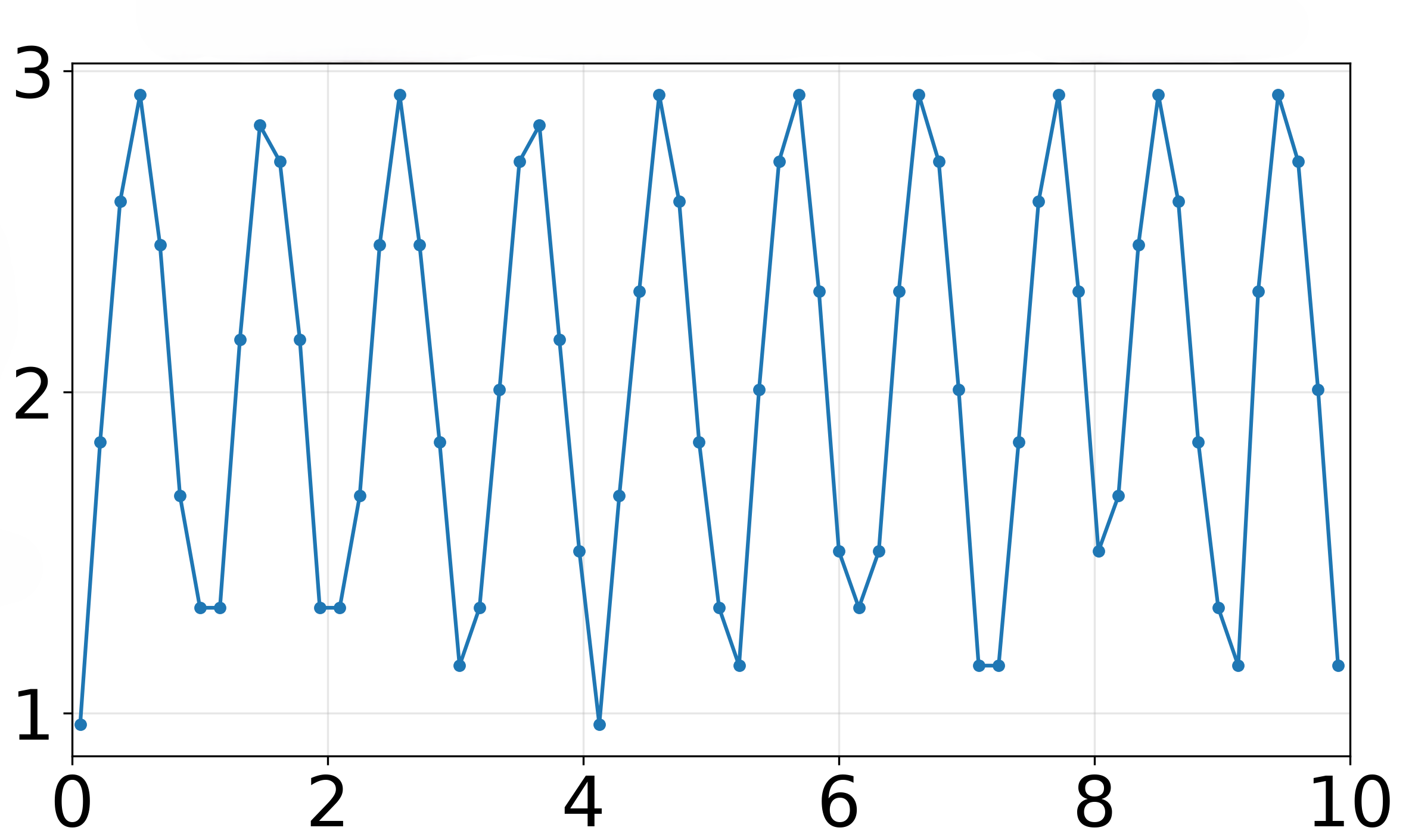}\\[-1mm]
        \normalsize Time (ms)
    }
    &
    \shortstack[c]{
        \hspace{2mm}
        \includegraphics[
            width=0.26\textwidth,
            height=0.19\textwidth
        ]{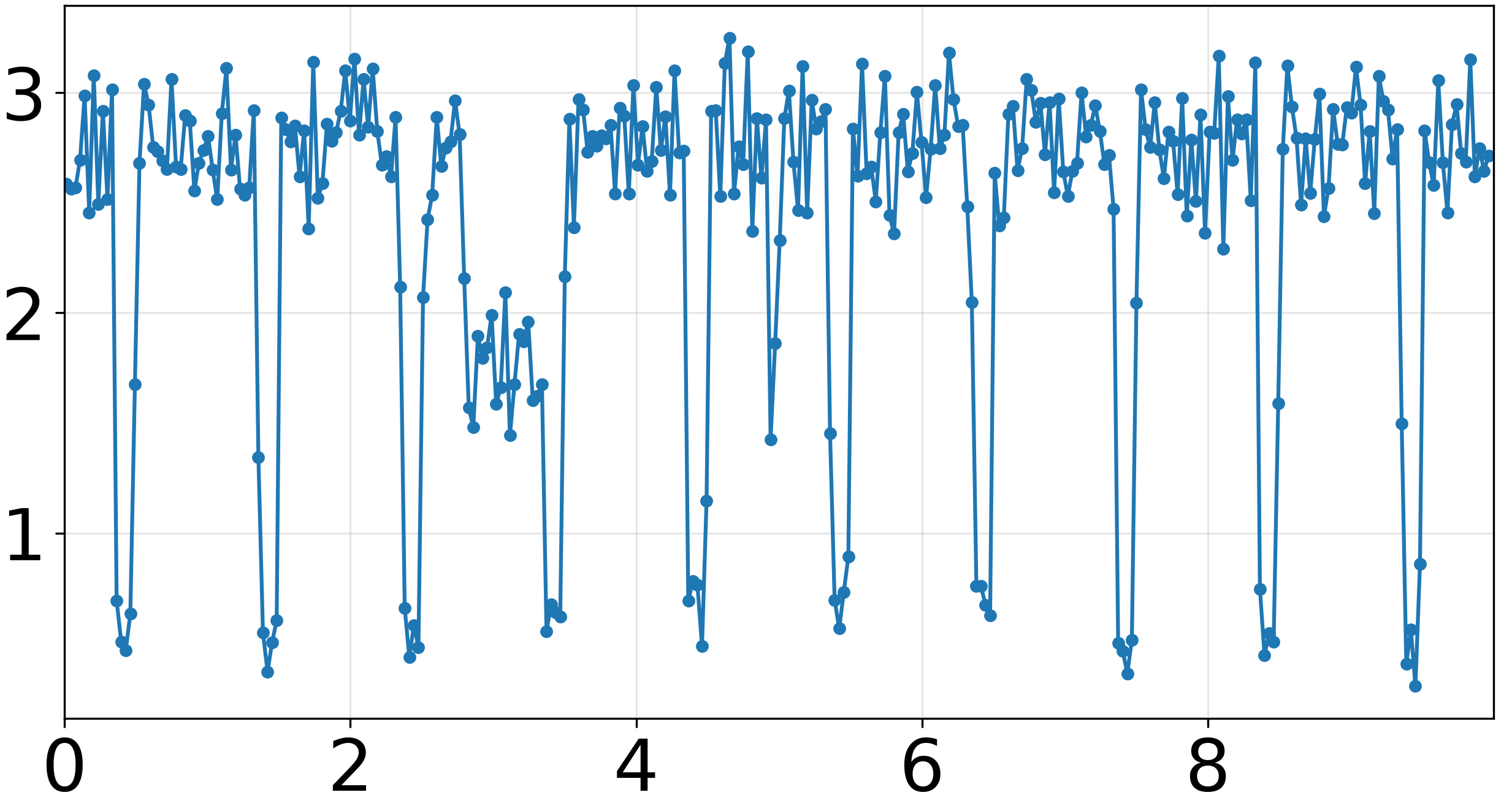}\\[-1mm]
        \normalsize Time (ms)
    }
    &
    \shortstack[c]{
        \hspace{-8mm}
        \includegraphics[
            width=0.28\textwidth,
            height=0.19\textwidth
        ]{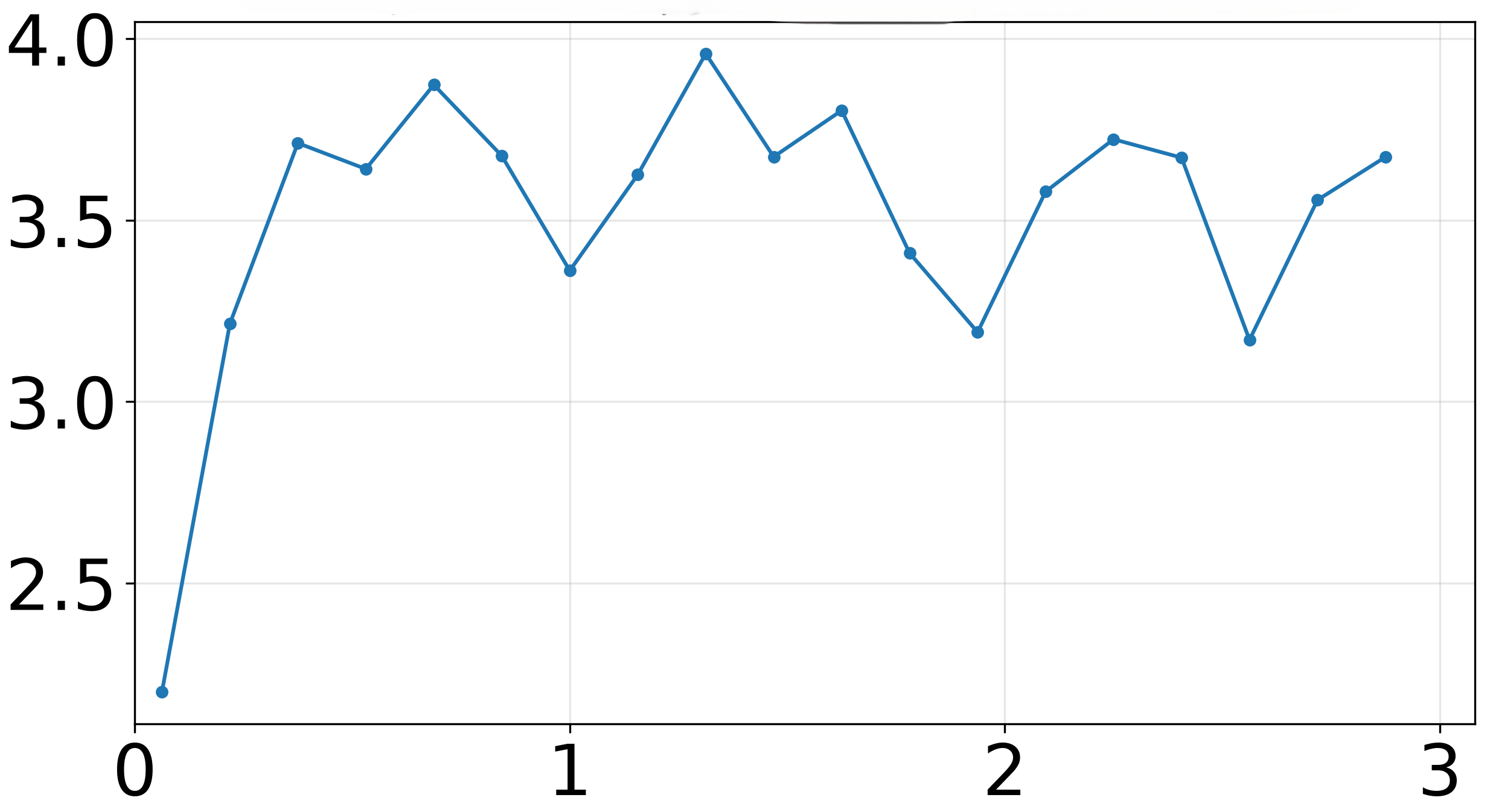}\\[-1mm]
        \normalsize Time (ms)
    }

    \end{tabular}

    \caption{Illustration of the proposed entropy representation for
    LoRa, 5G NR, and IEEE~802.11ah. The top row shows the corresponding
    spectrograms, while the bottom row presents the entropy sequences
    extracted from the same signals.}
    \label{fig:entropy_illustration}
\end{figure*}

By applying \eqref{eq:2d_dispersion_entropy} to consecutive time patches of
$\mathbf S_c$, an entropy sequence is obtained:
\[
\mathbf h =
\left[
h_1,h_2,\ldots,h_Q
\right]^T,
\]
where $Q$ is the number of retained entropy values. $Q$ is equal to the number of non-overlapping time-frequency patches extracted from the retained spectrogram and is given by

\begin{equation}
Q = \left\lfloor \frac{M}{M_p} \right\rfloor.
\end{equation}

Based on the simulation results, the combination $m_h=3$ and $m_w=2$ provided the best recognition performance among the evaluated embedding dimensions. This combination is also consistent with the embedding dimensions recommended for dispersion entropy in \cite{e23101303}.

The resulting entropy sequence $\mathbf h$ provides a compact description of
how the local time-frequency complexity evolves throughout the detected packet. Fig.~\ref{fig:entropy_illustration}
provides an example of this representation for three
different wireless technologies, namely LoRa, \ac{5g}, and IEEE~802.11ah. The first row shows the corresponding
spectrograms, while the second row presents the entropy sequences
obtained from the same spectrograms using the proposed 2D dispersion
entropy procedure.

The spectrograms exhibit different patterns in the distribution and
evolution of signal energy across the time-frequency plane. These
differences are reflected in the corresponding entropy sequences,
where the entropy values vary according to the local diversity of
dispersion patterns within each time-frequency patch. Thus, the
entropy sequence converts the two-dimensional local complexity of the
spectrogram into a compact one-dimensional representation that
describes how this complexity evolves over time. The different
entropy
trajectories observed for the three examples illustrate how
technology-dependent time-frequency characteristics can be reflected
in the resulting entropy representation.

\subsection{Statistical Feature Extraction}

The two-dimensional dispersion entropy stage produces an entropy sequence
$\mathbf{h}=[h_1,h_2,\ldots,h_Q]^T$. Since different wireless technologies may produce packets of different durations, resulting in entropy sequences of different lengths, directly using $\mathbf{h}$ as the classifier input would require either padding or truncation. Such operations may introduce artificial structure or discard informative entropy variations. Therefore, the proposed framework converts the variable-length entropy sequence into a fixed-dimensional statistical feature vector.

The feature vector is constructed by combining the initial entropy values with global statistical descriptors of the sequence. The
first four entropy samples are included explicitly as
\[
\mathbf h_{1:4}
=
[h_1,h_2,h_3,h_4]^T.
\]

These initial values are retained because the beginning of a packet
often contains technology-dependent structures such as preambles,
synchronization fields, or headers.

The remaining features summarize the distribution and temporal evolution of the entropy sequence. Specifically, the feature vector includes the mean $\mu_h$, median $\tilde{h}$, and standard deviation $\sigma_h$ of the entropy values, together with several percentile-based descriptors, namely the 10th $\mathcal Q_{10}$, 25th $\mathcal Q_{25}$, 75th $\mathcal Q_{75}$, and 90th $\mathcal Q_{90}$ percentiles, and the corresponding inter-quartile range $\mathrm{IQR}_h$. These features characterize the dispersion and the overall distribution of the entropy sequence.

Additional descriptors are included to capture the occurrence of highly regular time-frequency regions, the shape of the entropy distribution, and its robustness to outliers. These include the proportion of zero-valued entropy samples $r_0$, kurtosis $\kappa_h$, and the median absolute deviation $\mathrm{MAD}_h$.

To characterize the temporal evolution of the entropy sequence, first-order differences between consecutive entropy values are computed as
\begin{equation}
    d_q = h_{q+1} - h_q,\quad q=1,\ldots,Q-1.
\end{equation}

The mean $\mu_d$ and standard deviation $\sigma_d$ of the resulting difference sequence are included as features. These descriptors capture the average trend and variability of the entropy evolution along the packet.

Finally, three additional descriptors are extracted from the entropy sequence. The one-dimensional dispersion entropy $D_{1D}$ is computed from the entropy sequence itself and provides a measure of its complexity. Following the same principles as the 2D dispersion entropy, the entropy sequence is first quantized into a finite number of dispersion classes, from which dispersion patterns are formed and their probabilities estimated \cite{7434608}. The resulting one-dimensional dispersion entropy is denoted by $D_{1D}$. The uniqueness ratio $R_{\mathrm{uniq}}$ quantifies the diversity of the entropy sequence and is defined as \begin{equation} R_{\mathrm{uniq}}=\frac{|U(h)|}{Q}, \end{equation} where $U(h)$ denotes the set of unique values appearing in the entropy sequence. The collision probability $P_{\mathrm{coll}}$ measures the probability that two entropy samples drawn independently from the empirical distribution of the sequence have the same value. Let $p_j$ denote the empirical probability of the distinct entropy value $\nu_j$. The collision probability is defined as \begin{equation} P_{\mathrm{coll}} = \sum_{j=1}^{J} p_j^2, \end{equation} where $J=|U(h)|$ is the number of distinct entropy values. A high value of $P_{\mathrm{coll}}$ indicates that the entropy sequence is dominated by repeated values, whereas a low value indicates a more diverse entropy distribution.

The dimension of the final feature vector is selected based on a
sensitivity analysis of the feature dimension. To this end, an extended
feature vector containing additional candidate descriptors was initially
constructed, and feature subsets of increasing dimension were evaluated
using the same classification framework. The results showed that a
feature dimension of $d=20$ provided the best trade-off between
recognition performance and model size. Increasing the feature dimension
beyond $20$ did not provide a corresponding improvement in classification
performance while increasing the dimensionality of the classifier input
and, consequently, its model complexity. Therefore, $d=20$ is adopted
for the proposed feature representation.

The final feature vector is formed by concatenating the initial entropy values
and the statistical descriptors as
\begin{equation}
\mathbf x
=
\left[
\mathbf h_{1:4}^{T},
\mathbf s_h^{T}
\right]^T
\in \mathbb{R}^{20},
\label{eq:statistical_feature_vector}
\end{equation}
where $\mathbf s_h \in \mathbb{R}^{16}$ is defined as
\begin{align}
\mathbf s_h =
[
&\mu_h,\,
\tilde{h},\,
\sigma_h,\,
\mathrm{IQR}_h,\,
\mathcal Q_{25},\,
\mathcal Q_{75},\,
\mathcal Q_{10},\,
\mathcal Q_{90},
\nonumber\\
&r_0,\,
D_{\mathrm{1D}},\,
\kappa_h,\,
\mathrm{MAD}_h,\,
\mu_d,\,
\sigma_d,\,
R_{\mathrm{uniq}},\,
P_{\mathrm{coll}}
]^T.
\label{eq:statistical_descriptor_vector}
\end{align}

The resulting fixed-length representation $\mathbf x$
summarizes the level, spread, shape, diversity, and temporal variation of the
entropy sequence and is used as the input to the wireless technology
classifier.

\subsection{MLP-Based Wireless Technology Classification}

The fixed-dimensional statistical feature vector
$\mathbf x \in \mathbb{R}^{20}$ obtained from the entropy sequence is used as
the input to a lightweight \ac{mlp} classifier.

Before classification, each feature is standardized using the mean and
standard deviation computed from the training set. Let $\mu_i$ and $\sigma_i$
denote the training-set mean and standard deviation of the $i$th feature,
respectively. The normalized feature vector is given by
\begin{equation}
\hat{x}_i =
\frac{x_i-\mu_i}{\sigma_i+\epsilon},
\qquad i=1,\ldots,20,
\label{eq:feature_standardization}
\end{equation}
where $\epsilon$ is a small positive constant used for numerical stability.

The \ac{mlp} maps the normalized feature vector $\hat{\mathbf x}$ to a vector of
class logits. The classifier consists of two hidden layers followed by an
output layer. Each hidden layer applies a batch
normalization, a \ac{relu} activation, and dropout
regularization. The network can be written as
\begin{align}
\mathbf u_1
&=
\operatorname{Dropout}
\left(
\operatorname{ReLU}
\left(
\operatorname{BN}
\left(
\mathbf W_1 \hat{\mathbf x}+\mathbf b_1
\right)
\right)
\right),
\label{eq:mlp_layer1}
\\
\mathbf u_2
&=
\operatorname{Dropout}
\left(
\operatorname{ReLU}
\left(
\operatorname{BN}
\left(
\mathbf W_2 \mathbf u_1+\mathbf b_2
\right)
\right)
\right),
\label{eq:mlp_layer2}
\\
\mathbf z
&=
\mathbf W_3 \mathbf u_2+\mathbf b_3,
\label{eq:mlp_logits}
\end{align}
where $\mathbf W_i$ and $\mathbf b_i$ are the trainable weights and biases,
$\operatorname{BN}(\cdot)$ denotes batch normalization, and
$\mathbf z \in \mathbb{R}^{C_{\mathrm{cls}}}$ is the output logit vector for
$C_{\mathrm{cls}}$ wireless technology classes.

In this work, the MLP architecture is
\[
20 \rightarrow 128 \rightarrow 64 \rightarrow C_{\mathrm{cls}},
\]
where the input dimension is determined by the 20 extracted statistical
features. Dropout with probability $0.3$ is applied after each hidden layer to
reduce overfitting.

The posterior probability assigned to class $c$ is obtained using the softmax
function:
\begin{equation}
p(c|\mathbf x)
=
\frac{
\exp(z_c)
}{
\sum_{j=1}^{C_{\mathrm{cls}}}
\exp(z_j)
},
\qquad
c=1,\ldots,C_{\mathrm{cls}}.
\label{eq:softmax_probability}
\end{equation}

The predicted wireless technology is therefore
\begin{equation}
\hat{c}
=
\arg\max_{c}
p(c|\mathbf x).
\label{eq:predicted_class}
\end{equation}

The network is trained by minimizing the weighted cross-entropy loss. For a
training set of $N_{\mathrm{tr}}$ feature-label pairs
$\{(\mathbf x_i,y_i)\}_{i=1}^{N_{\mathrm{tr}}}$, the loss is defined as
\begin{equation}
\mathcal L
=
-
\frac{1}{N_{\mathrm{tr}}}
\sum_{i=1}^{N_{\mathrm{tr}}}
\alpha_{y_i}
\log
p(y_i|\mathbf x_i),
\label{eq:weighted_cross_entropy}
\end{equation}
where $y_i$ is the true class label of the $i$th training sample and
$\alpha_{y_i}$ is the weight assigned to its class. The use of class weights
compensates for possible class imbalance in the training data. The \ac{mlp}
parameters are optimized using the Adam optimizer with weight decay
regularization. Early stopping is applied based on the validation loss to avoid
overfitting.

The output of this stage is the estimated wireless technology label associated
with the detected packet. Since the classifier operates only on the compact
statistical representation $\mathbf x$, the overall recognition pipeline
remains computationally efficient while preserving the discriminative
information extracted from the two-dimensional entropy sequence.

\begin{table*}[!t]
\caption{IQ-Based and STFT-Based CNN Models Architectures}
\label{tab:cnn_architecture}
\centering
\footnotesize
\renewcommand{\arraystretch}{1.2}
\begin{tabular}{|l|p{5.3cm}|p{5.3cm}|}
\hline
\textbf{Layer} & \textbf{IQ-Based CNN} & \textbf{STFT-Based CNN} \\
\hline
Input
&
$1\times2\times1024$ IQ samples
&
$1\times F\times T$ magnitude spectrogram
\\
\hline
Conv Block 1
&
Conv$(1\times3)$, 64 filters, BatchNorm, ReLU, MaxPool$(1\times2)$, Dropout(0.3)
&
Conv$(3\times3)$, 32 filters, BatchNorm, ReLU, MaxPool$(2\times2)$
\\
\hline
Conv Block 2
&
Conv$(2\times3)$, 32 filters, BatchNorm, ReLU, MaxPool$(1\times2)$, Dropout(0.3)
&
Conv$(3\times3)$, 64 filters, BatchNorm, ReLU, MaxPool$(2\times2)$
\\
\hline
Conv Block 3
&
Conv$(1\times3)$, 16 filters, BatchNorm, ReLU, MaxPool$(1\times2)$, Dropout(0.3)
&
Conv$(3\times3)$, 128 filters, BatchNorm, ReLU, MaxPool$(2\times2)$
\\
\hline
Conv Block 4
&
Not used
&
Conv$(3\times3)$, 256 filters, BatchNorm, ReLU
\\
\hline
Pooling
&
MaxPooling after each convolutional block
&
Adaptive Global Average Pooling $(1\times1)$
\\
\hline
Fully Connected
&
Flatten $\rightarrow$ 1000, ReLU, Dropout(0.3)
&
256 $\rightarrow$ 128, ReLU, Dropout(0.3)
\\
\hline
Output
&
1000 $\rightarrow$ 7 classes
&
128 $\rightarrow$ 7 classes
\\
\hline
\end{tabular}
\end{table*}

\subsection{Training Configurations}
\label{subsec:TrainingConfigurations}

In practical \ac{wtr} systems, the operating
conditions may vary considerably with the received signal quality.
Training with noisy observations can improve the ability of a
classifier to recognize signals under low-\ac{snr} conditions by exposing
the model to channel conditions similar to those encountered during
operation. To investigate the effect of such training,
two training configurations are considered in this work:

\begin{enumerate}
    \item \textbf{Unaugmented-training configuration:} The training set consists only of packets extracted from the clean recordings, whereas the test set contains the corresponding packets from all considered \ac{snr} levels.

    \item \textbf{Augmented-training configuration:} The training set is expanded by including the noisy realizations of the training packets for all considered \ac{snr} levels, while the test set remains identical to that of the unaugmented setting.
\end{enumerate}

The packet partition is established using the clean recordings and is preserved for all noisy realizations. Consequently, every packet and all of its noisy realizations belong exclusively to either the training or testing set, thereby preventing data leakage and ensuring a fair performance evaluation.

\subsection{Benchmark Methods}

To provide a reference for performance assessment, two \ac{dl} baseline models are implemented in addition to the proposed entropy-based framework: an \ac{iq}-based \ac{cnn} and a \ac{stft}-based \ac{cnn}. Both models use the same packet-detection procedure, dataset partitioning strategy, and training configurations as the proposed method to ensure a consistent comparison.

For the \ac{iq}-based \ac{cnn}, each detected packet is segmented into non-overlapping windows containing 1024 \ac{iq} samples. The in-phase and quadrature components are used as two input channels and are provided directly to the network. The network architecture is inspired by the \ac{cnn} framework proposed by Merkebu Girmay et al.~\cite{GIRMAY2023100563} for \ac{wtr}, with some modifications to accommodate the raw \ac{iq} input representation.

For the \ac{stft}-based \ac{cnn}, the same 1024-sample \ac{iq} windows are transformed into magnitude spectrograms. Before classification, the spectrograms undergo the same pre-processing stages used by the proposed framework, including \ac{cfo} compensation and spectrogram thresholding. The processed spectrograms are subsequently provided to a separate two-dimensional \ac{cnn} designed to exploit the time-frequency structure of the spectrogram representation. The architecture and its hyper-parameters were selected experimentally based on recognition performance.

Both \acp{cnn} are trained using the Adam optimizer with a learning rate of $10^{-3}$ and a weight decay of $10^{-4}$. The cross-entropy loss function is employed with class weighting to compensate the class imbalance. The models are trained using a batch size of 128 with early stopping based on the validation loss. Table~\ref{tab:cnn_architecture} summarizes the architectures of the two \ac{cnn} benchmark models.

\section{Results and Discussion}

This section evaluates the effectiveness of the proposed entropy-based \ac{wtr} framework from both performance and computational perspectives. Recognition accuracy is first examined under different \ac{snr} conditions and compared with two \ac{cnn}-based baselines. Next, the effect of including noisy samples in the training on the recognition performance of all methods is investigated. Finally, the computational complexity of the approaches is analyzed to highlight the trade-offs between recognition performance, training cost, and inference cost.

\subsection{Recognition Performance under Unaugmented Training}

\begin{figure}[!t]
    \centering
    \includegraphics[width=\columnwidth]{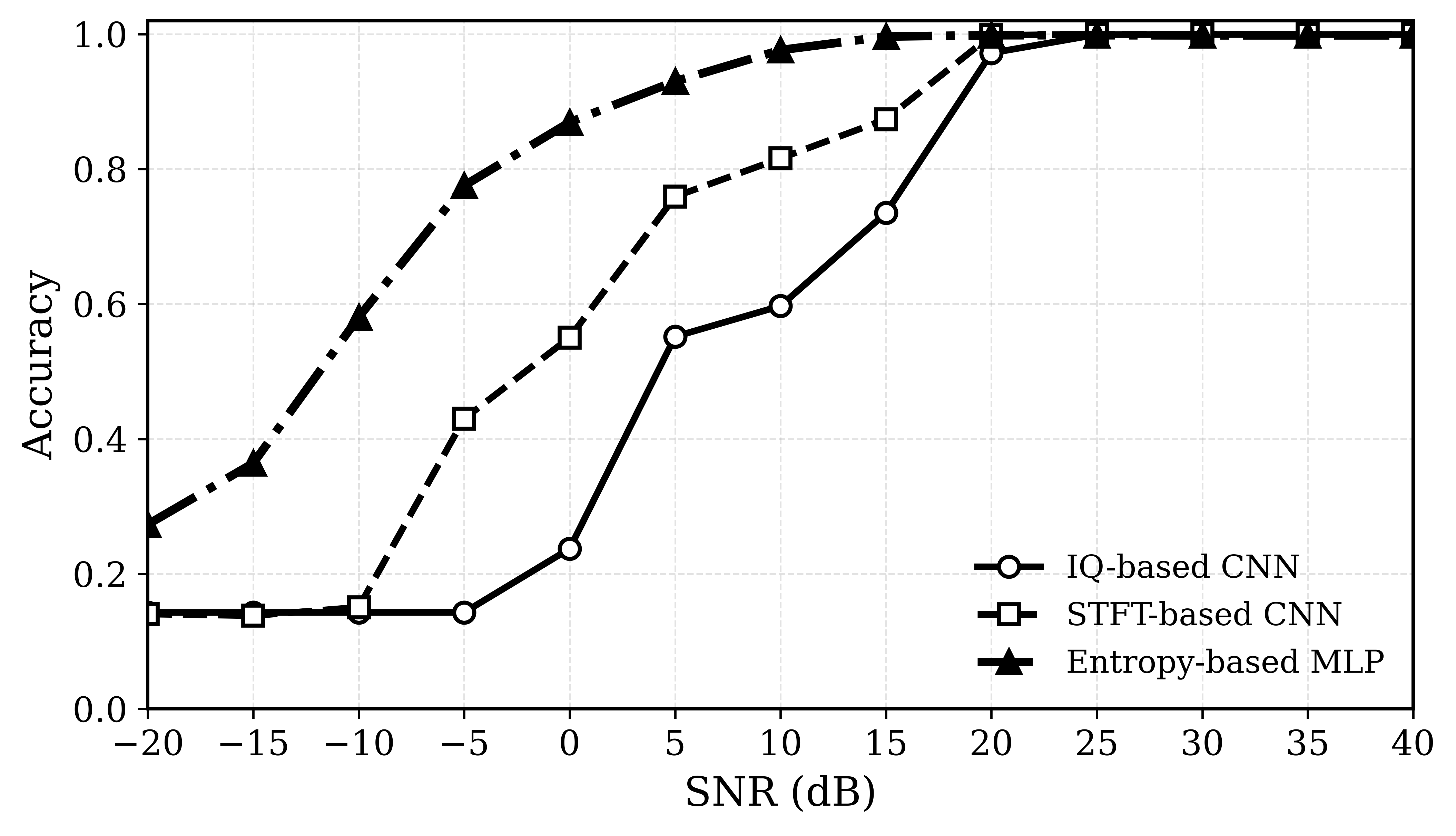}
    \caption{Recognition accuracy vs. \ac{snr} under unaugmented-training
configuration}
    \label{fig:AccuracyWithoutAugmentation}
\end{figure}

\begin{figure*}[!t]
    \centering 
    \includegraphics[width=\textwidth]{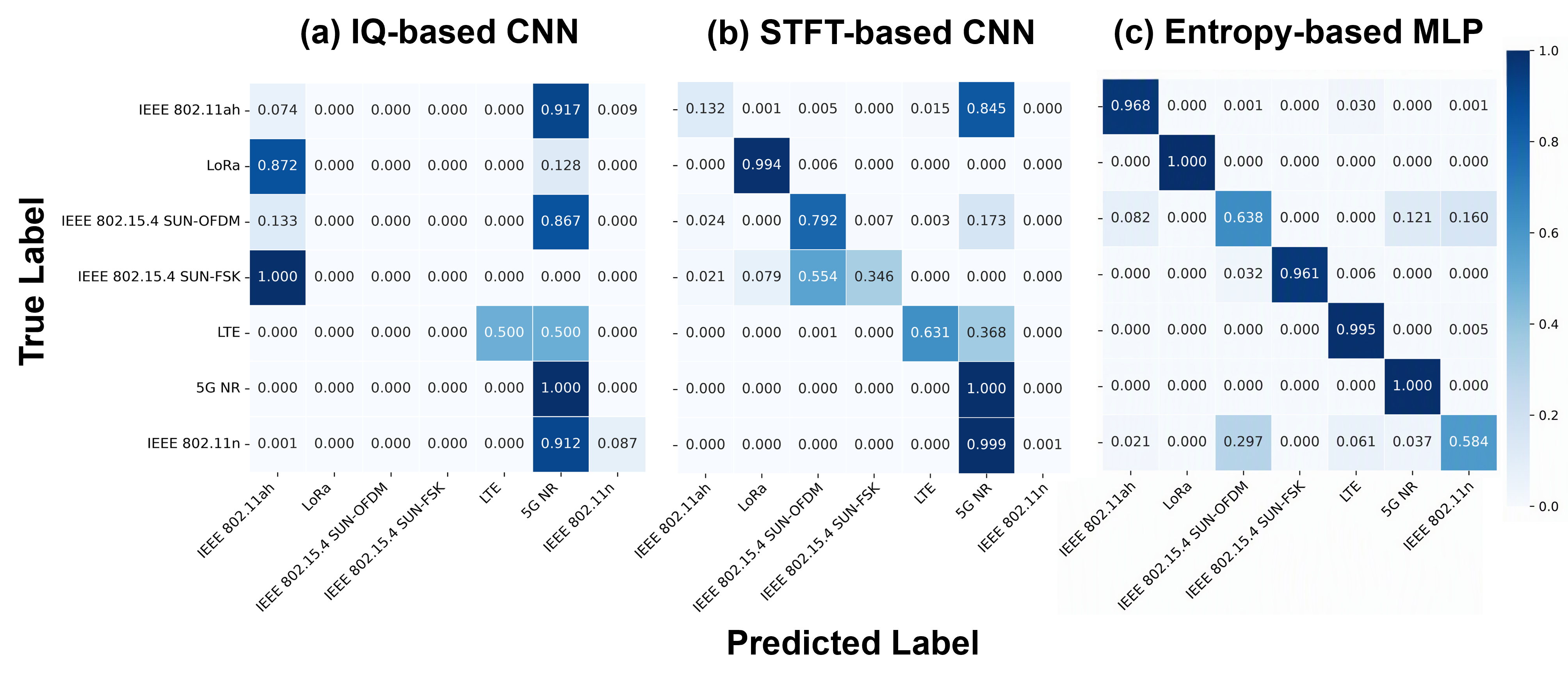} 
    \caption{Confusion matrix at \ac{snr} of 0 dB under the unaugmented-training configuration} 
    \label{fig:un-0} 
\end{figure*} 

This subsection evaluates the recognition performance of the proposed framework and the \ac{cnn}-based baselines under the unaugmented training strategy described in Subsection~\ref{subsec:TrainingConfigurations}.

As shown in Fig.~\ref{fig:AccuracyWithoutAugmentation}, the proposed entropy-based framework consistently outperforms the \ac{cnn}-based approaches over the low- and moderate-\ac{snr} regions. The performance advantage is particularly pronounced under challenging channel conditions, where the entropy-based feature representation preserves its discriminative capability despite the presence of severe noise. Averaged over all evaluated \ac{snr} levels, the proposed framework achieves a recognition accuracy of 82.73\%, compared with 68.08\% for the \ac{stft}-based \ac{cnn} and 58.93\% for the \ac{iq}-based \ac{cnn}, demonstrating an overall improvement.

As the \ac{snr} increases, the recognition performance of all methods improves and gradually converges. For \ac{snr} values of 20~dB and above, all approaches achieve nearly perfect recognition accuracy, indicating that the received signals contain sufficient discriminative information for reliable classification. The primary distinction between the compared methods therefore lies in their robustness under adverse channel conditions, where the proposed entropy-based framework demonstrates a clear advantage.

The \ac{stft}-based \ac{cnn} exhibits a clear advantage over the \ac{iq}-based \ac{cnn} at low \ac{snr}. This can be attributed to the structured time-frequency
representation provided by the \ac{stft}, which makes spectral and
temporal characteristics of the received signal more explicit to the
\ac{cnn}. This observation is consistent with the findings reported
in~\cite{FONTAINE2019101881}, where frequency-domain spectrogram
representations generated from \ac{iq} samples were shown to maintain
high recognition accuracy even at low \ac{snr}. In contrast, the
\ac{iq}-based \ac{cnn} learns the relevant signal characteristics
directly from the raw samples and is therefore more sensitive to the
presence of noise.

Fig.~\ref{fig:un-0} provides a detailed view of the class-wise recognition performance of the three considered approaches at \ac{snr} of $0$~dB. A clear difference can be observed in the
discriminative ability of the representations provided to the classifiers. The \ac{iq}-based \ac{cnn} exhibits substantial
inter-class confusion, with several technologies being incorrectly
assigned to other classes. The \ac{stft}-based \ac{cnn} provides a
considerable improvement by exposing time-frequency characteristics
that are more informative for technology recognition; however,
significant confusion remains for several classes, such as IEEE~802.11ah, IEEE~802.15.4~SUN-FSK, LTE, and IEEE~802.11n.

In contrast, the proposed entropy-based representation produces a
substantially more discriminative class structure. The corresponding
confusion matrix shows strong diagonal dominance for most of the
considered technologies.
In particular, LoRa and 5G NR are completely separated at this
\ac{snr}, while IEEE~802.11ah, IEEE~802.15.4~SUN-FSK, and LTE also
exhibit very high class-wise recognition rates. Although some
confusion remains between IEEE~802.15.4~SUN-OFDM and IEEE~802.11n,
the entropy-based representation substantially reduces the
inter-class confusion observed with the \ac{iq}- and \ac{stft}-based
representations.

This behavior highlights the discriminative capability of the
proposed representation. Rather than relying directly on the detailed
amplitude structure of the received \ac{iq} samples or on the
time-frequency energy distribution alone, the proposed method
characterizes the local complexity and its temporal evolution in the
spectrogram. The resulting entropy features therefore expose
technology-dependent characteristics that remain sufficiently
distinct for classification even under the noisy $0$~dB testing
condition. The substantially stronger diagonal dominance of the
entropy-based confusion matrix demonstrates that this representation
provides a more separable feature space for \ac{wtr} under the considered unaugmented training
configuration.

\subsection{Recognition Performance under Augmented Training}

\begin{figure}[!t]
    \centering
    \includegraphics[width=\columnwidth]{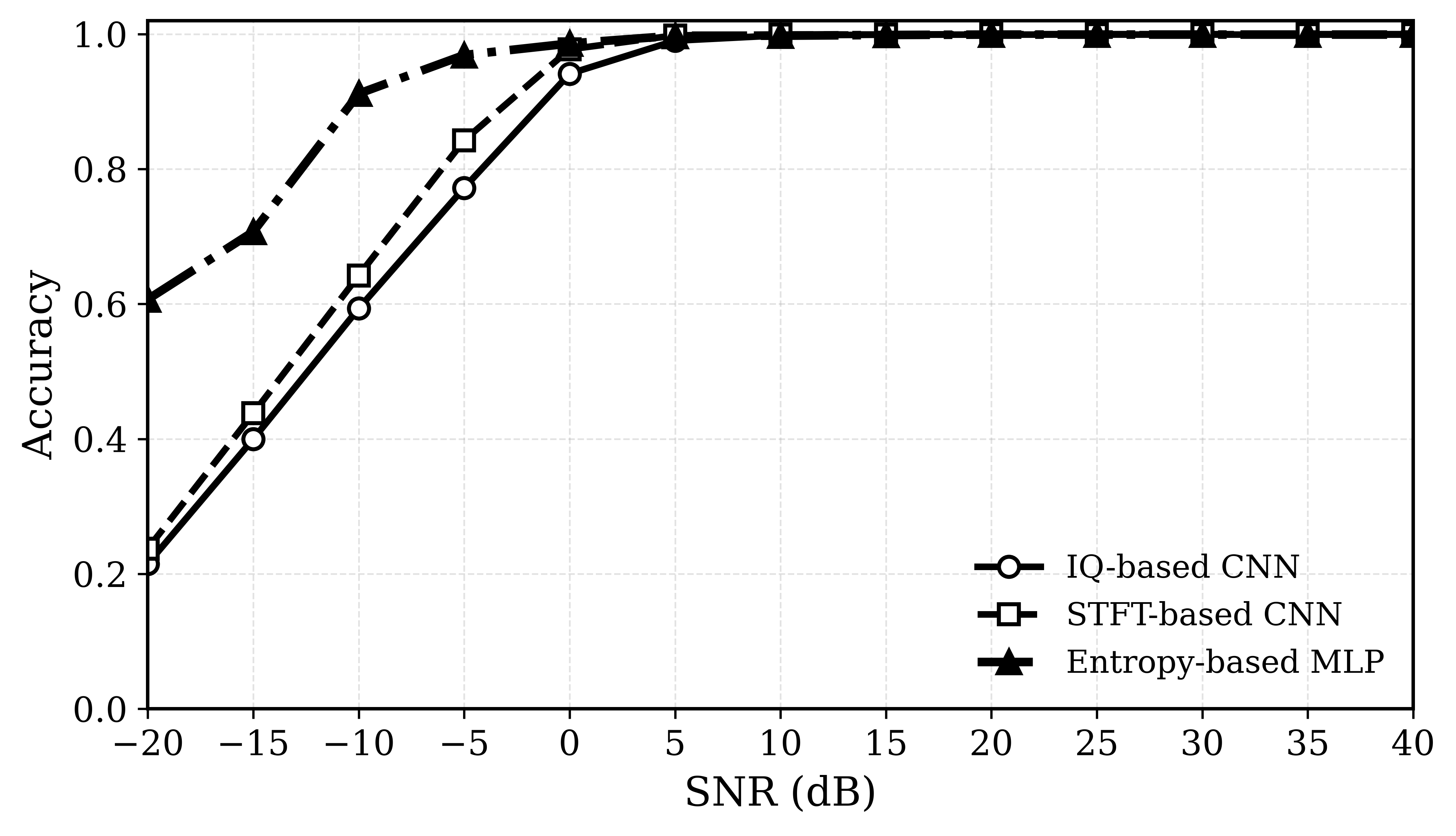}
    \caption{Recognition accuracy vs. \ac{snr} under augmented-training
configuration}
    \label{fig:AccuracyWithAugmentation}
\end{figure}

Although the results presented in the previous subsection demonstrate the superior robustness of the proposed framework when trained exclusively using clean recordings, practical \ac{ml} systems often employ data augmentation to improve their robustness to noisy operating conditions. In particular, exposing the classifier to noisy training samples may improve the performance at low \ac{snr} values. To assess the effectiveness of this strategy, all methods are additionally evaluated using the augmented training setting described in Subsection~\ref{subsec:TrainingConfigurations}.

\begin{figure*}[!ht]
    \centering 
    \includegraphics[width=\textwidth]{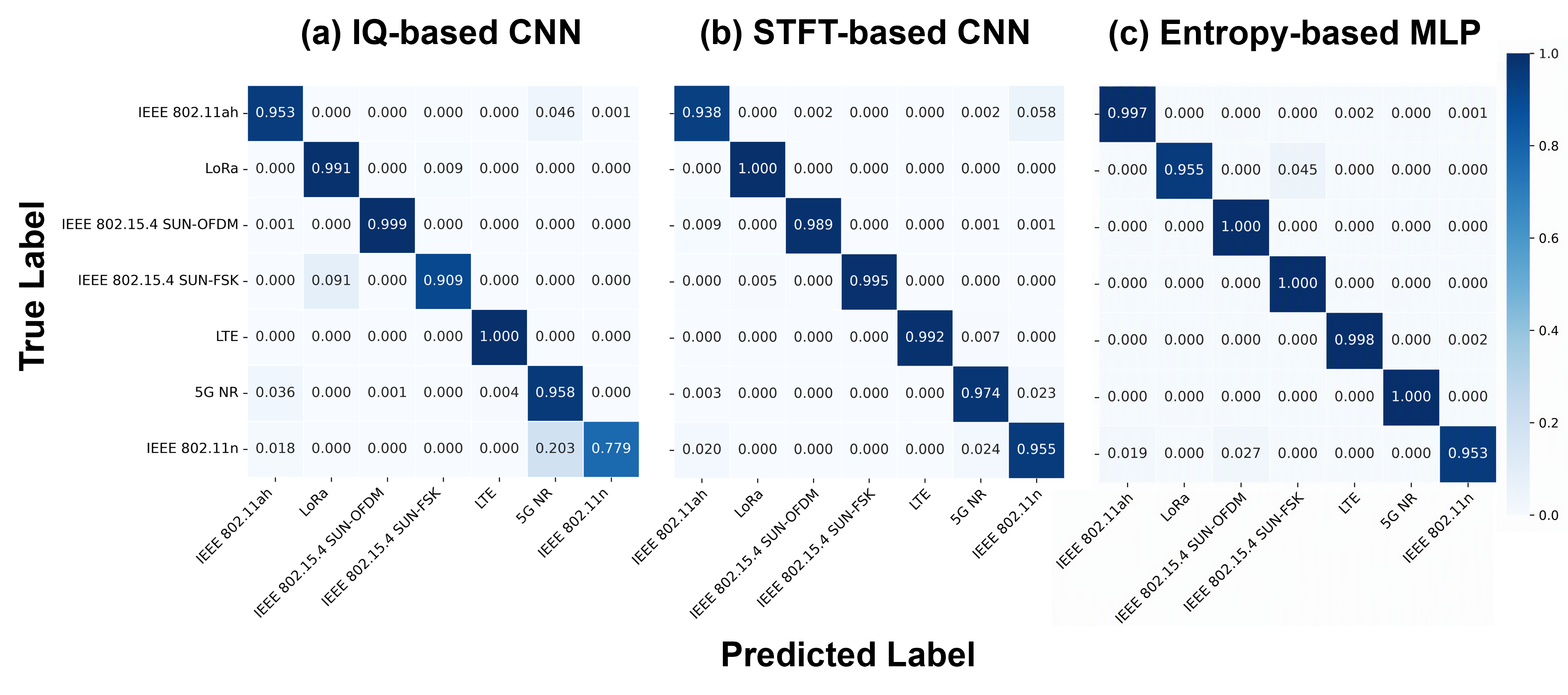} 
    \caption{Confusion matrix at \ac{snr} of 0 dB under the augmented-training configuration} 
    \label{fig:au-0} 
\end{figure*} 

\begin{table*}[t] 
\caption{Training Computational Performance} 
\label{tab:TrainingPerformance} 
\centering 
\begin{tabular}{lcccccc} 
\hline 
Method & Unit & Pre-processing/Unit (ms) & No. Units & Training/Epoch (s) & No. Epochs & Total Training (s) \\
\hline 
IQ-CNN 
& Window 
& 0.688
& 158965 
& 25.43 
& 38
& 1075.708 \\ 
STFT-CNN 
& Window
& 0.917 
& 158965 
& 36.722
& 13
& 623.157 \\ 
Proposed 
& Packet 
& 8.02 
& 21435 
& 0.745
& 20 
& 186.809 \\ 
\hline 
\end{tabular} 
\end{table*}

\begin{table*}[t] 
\caption{Inference Computational Performance} 
\label{tab:InferencePerformance} 
\centering 
\begin{tabular}{lccccc} 
\hline 
Method & Unit & Pre-processing/Unit (ms) & Inference/Unit ($\mu$s) & No. Units & Total Inference (s) \\
\hline 
IQ-CNN 
& Window 
& 0.688
& 131.57
& 2021799  
& 1657.006 \\ 
STFT-CNN 
& Window 
& 0.917
& 111.15 
& 2021799
& 2078.713 \\ 
Proposed 
& Packet 
& 8.02
& 30.09 
& 278499
& 2241.942 \\ 
\hline 
\end{tabular} 
\end{table*}

Fig.~\ref{fig:AccuracyWithAugmentation} shows the recognition accuracy versus \ac{snr} obtained under the augmented training strategy. As expected, all methods benefit from the inclusion of noisy training samples, particularly in the low- and moderate-\ac{snr} regions. Nevertheless, the proposed framework maintains the highest overall recognition performance, demonstrating that its entropy-based feature representation remains highly discriminative even when the competing \ac{cnn}-based approaches are trained using diverse data.

A comparison between Figs.~\ref{fig:AccuracyWithoutAugmentation}
and~\ref{fig:AccuracyWithAugmentation} further shows that the
performance gap between the \ac{iq}-based and \ac{stft}-based
\acp{cnn} becomes considerably smaller when noisy samples are
included during training. When noisy samples are included during training, the \ac{iq}-based \ac{cnn} is exposed to the signal variations that were absent from the
clean-training configuration and can consequently learn better noise-robust representations directly from the raw \ac{iq} samples. This results in a substantial improvement at low \ac{snr}. The
\ac{stft}-based \ac{cnn} also benefits from the augmented training;
however, its improvement is comparatively smaller, as the
time-frequency representation already provides a structured
representation of the signal characteristics relevant for
recognition. As a result, noisy-sample training reduces the
performance difference between the two \ac{cnn}-based approaches.

To further examine the class-wise recognition behavior under the
augmented-training configuration, Fig.~\ref{fig:au-0} presents the
confusion matrices of the \ac{iq}-based \ac{cnn}, \ac{stft}-based
\ac{cnn}, and the proposed entropy-based \ac{mlp} at \ac{snr} of
$0$~dB. In contrast to the unaugmented configuration, all three
approaches exhibit substantially stronger diagonal dominance, showing
that exposure to noisy training samples improves the ability of the
learned representations to distinguish the considered wireless
technologies under noisy conditions.

The three approaches nevertheless exhibit different class-wise
confusion patterns. For the \ac{iq}-based \ac{cnn}, most classes are
recognized with high accuracy, but IEEE~802.11n remains the most
challenging class, with $77.9\%$ of its samples correctly classified.
In particular, $20.3\%$ of IEEE~802.11n samples are classified as
\ac{5g}, while IEEE~802.15.4~SUN-FSK also exhibits some confusion with
LoRa, with $9.1\%$ of its samples assigned to that class. The
\ac{stft}-based \ac{cnn} provides improved class separation, with
recognition rates ranging from $93.8\%$ to $100\%$. Its remaining
confusions are comparatively small, with the largest occurring for
IEEE~802.11ah, for which $5.8\%$ of the samples are classified as
IEEE~802.11n.

The proposed entropy-based representation exhibits the strongest
class-wise separation among the three approaches. Its confusion matrix
is almost perfectly diagonal, with recognition rates of $99.7\%$,
$95.5\%$, $100\%$, $100\%$, $99.8\%$, $100\%$, and $95.3\%$ for
IEEE~802.11ah, LoRa, IEEE~802.15.4~SUN-OFDM,
IEEE~802.15.4~SUN-FSK, LTE, 5G NR, and IEEE~802.11n, respectively.
The largest off-diagonal entry is only $4.5\%$, corresponding to LoRa
samples classified as IEEE~802.15.4~SUN-FSK. The remaining
misclassifications are limited to a few percent.

These results highlight the discriminative capability of the proposed
entropy-based representation. Although all three approaches benefit
from augmented training, the entropy representation yields a more
distinct class structure at the feature level, resulting in fewer
inter-class confusions after classification. By characterizing the
local complexity and temporal evolution of the time-frequency
representation, the proposed features capture technology-dependent
characteristics that remain distinguishable even at an \ac{snr} of
$0$~dB. The near-diagonal confusion matrix therefore provides direct
evidence that the proposed entropy representation offers a highly
discriminative representation for \ac{wtr} under noisy conditions.

\subsection{Computational Performance}

The computational performance of the evaluated methods is assessed in terms of pre-processing, training, and inference cost. All experiments were conducted using JupyterLab on the GPULab computing infrastructure\footnote{GPULab: \url{https://idlab.ugent.be/resources/gpulab}}, using a workstation equipped with two Intel\textsuperscript{\textregistered} Xeon\textsuperscript{\textregistered} Silver 4310 processors (24 physical cores and 48 threads in total) and an NVIDIA A40 GPU with 46~GB of memory. 

Tables~\ref{tab:TrainingPerformance} and~\ref{tab:InferencePerformance} summarize the computational performance of the evaluated methods. Due to their different input representations (Unit), the \ac{cnn}-based approaches process fixed-length windows of 1024 \ac{iq} samples, whereas the proposed framework operates on complete detected packets. 

Table~\ref{tab:TrainingPerformance} shows that the proposed entropy-based framework achieves the lowest overall training cost. Owing to the compact 20-dimensional feature representation and lightweight \ac{mlp} classifier, each training epoch is substantially faster than for the \ac{cnn}-based baselines, resulting in a total training time of 186.8~s compared with 623.2~s and 1075.7~s for the \ac{stft}-based and \ac{iq}-based \acp{cnn}, respectively. 

Table~\ref{tab:InferencePerformance} presents the inference-related computational costs. While the proposed approach shifts a larger fraction of the computational burden to the entropy extraction stage, the subsequent classification stage remains highly efficient. 

Overall, the proposed framework provides an attractive trade-off between recognition performance and computational complexity. In particular, it achieves superior recognition performance while substantially reducing training complexity and maintaining practical inference times.

\section{Conclusion}

This paper investigated whether wireless technologies can be distinguished through the temporal evolution of their local time-frequency complexity. To capture this behavior, we proposed a lightweight \ac{wtr} framework based on two-dimensional dispersion entropy, which extracts compact technology-specific fingerprints from localized spectrogram regions and classifies them using an \ac{mlp}. Experimental results on seven wireless technologies demonstrated that the proposed entropy-based representation consistently outperforms \ac{iq}-based and \ac{stft}-based \ac{cnn} baselines, particularly under low- and moderate-\ac{snr} conditions. Furthermore, the proposed approach maintained its recognition advantage under data-augmented training while requiring substantially shorter training times than the \ac{cnn}-based approaches. The results indicate that the temporal evolution of local spectrogram complexity provides a robust technology-specific fingerprint that remains discriminative even under severe noise conditions. These findings suggest that entropy-based representations constitute a promising alternative to both traditional handcrafted descriptors and computationally intensive \ac{dl} approaches for \ac{wtr}.

\bibliographystyle{IEEEtran}
\bibliography{bibliography/references}

\end{document}